%% file: acl_latex.tex
\documentclass[11pt]{article}
\usepackage{acl}
\usepackage{times}
\usepackage{latexsym}
\usepackage{amsmath,amssymb}
\usepackage[T1]{fontenc}
\usepackage[utf8]{inputenc}
\usepackage{microtype,inconsolata}
\usepackage{graphicx,booktabs,xcolor,multirow}
\usepackage{makecell}

\usepackage{cuted}
\usepackage{capt-of}

\definecolor{nllgreen}{RGB}{0,120,70}
\definecolor{nllred}{RGB}{185,45,45}
\newcommand{\llgain}[1]{\textcolor{nllgreen}{\ensuremath{#1}}}
\newcommand{\llloss}[1]{\textcolor{nllred}{\ensuremath{#1}}}
\newcommand{\nllsupport}[2]{\ensuremath{#1\!\rightarrow\!\textcolor{nllgreen}{#2}}}
\newcommand{\nllsuppress}[2]{\ensuremath{#1\!\rightarrow\!\textcolor{nllred}{#2}}}

\title{Does Expert Supervision Improve Text-to-Speech?\\An Empirical Study of Preference Optimization}
\author{
    Minh Duc Vu$^{1}$\thanks{These authors contributed equally.}
    \quad
    Junchuan Zhao$^{2}$\footnotemark[1]
    \quad
    Ye Wang$^{2}$\thanks{Corresponding author.}
    \\
    $^{1}$Department of Statistics \& Data Science, National University of Singapore, Singapore \\
    $^{2}$School of Computing, National University of Singapore, Singapore \\
    \texttt{minhduc.vu@u.nus.edu},
    \texttt{junchuan@u.nus.edu},
    \texttt{dcswangy@nus.edu.sg}
}

\begin{document}
\maketitle
\begin{abstract}
\input{Sections/abstract}
\end{abstract}

\section{Introduction}
\label{sec:intro}
\input{Sections/intro}

\section{Related Work}
\label{sec:relate}
\input{Sections/related}

\section{Study Design}
\label{sec:study-design}
\input{Sections/study_design}

\section{Experimental Setup}
\label{sec:exp-settings}
\input{Sections/experimental_setup}

\section{Results}
\label{sec:results}
\input{Sections/results}

\section{Conclusion}
\label{sec:conclusion}
\input{Sections/conclusion}

\clearpage
\section*{Limitations}
\input{Sections/limitations}

\begingroup
\emergencystretch=1em
\bibliography{custom}
\endgroup

\appendix

\input{Appendix/appendix}

\end{document}

%% file: Sections/abstract.tex
Preference optimization promises to improve text-to-speech (TTS) systems by learning which speech outputs are better. External expert speech seems a natural source of preferred responses, especially when student generations contain errors. Yet demonstration quality does not establish whether the student can benefit from that supervision, or whether better target fitting will produce better speech. We examine this gap across two TTS students using real and synthetic expert speech, comparing preference objectives and direct versus bridge supervision while tracking fixed-sequence likelihoods and generation metrics. With expert supervision held fixed, objectives differ markedly in their sensitivity to preference-score scaling and the resulting generation quality. Likelihood diagnostics characterize these responses, but cannot establish their benefit: expert likelihood can increase while intelligibility worsens, or decrease while it improves. Bridge supervision highlights a distinction between mitigating degradation and improving the student: it can recover performance lost under direct training without consistently surpassing the pretrained base. Route benefits vary across objectives, students, and expert sources; neither route consistently dominates. Together, these findings distinguish demonstration quality, target fitting, and generated-speech quality as complementary criteria for assessing expert supervision.

%% file: Sections/intro.tex
Preference optimization offers text-to-speech (TTS) systems a way to learn which speech outputs to prefer. Language-model-based TTS enables this approach through speech-token prediction \citep{chen2025valle,peng2024voicecraft,casanova2024xtts,du2024cosyvoice,anastassiou2024seedtts,hu2026qwen3tts}. Direct Preference Optimization (DPO) learns from chosen--rejected pairs without a separately trained reward model \citep{rafailov2023dpo}, and has been applied to speech generation \citep{zhang2024speechalign,zhang2025intp}. Its appeal, however, depends on where those preferred responses come from.

\begin{figure}[t]
\centering
\includegraphics[width=\linewidth]{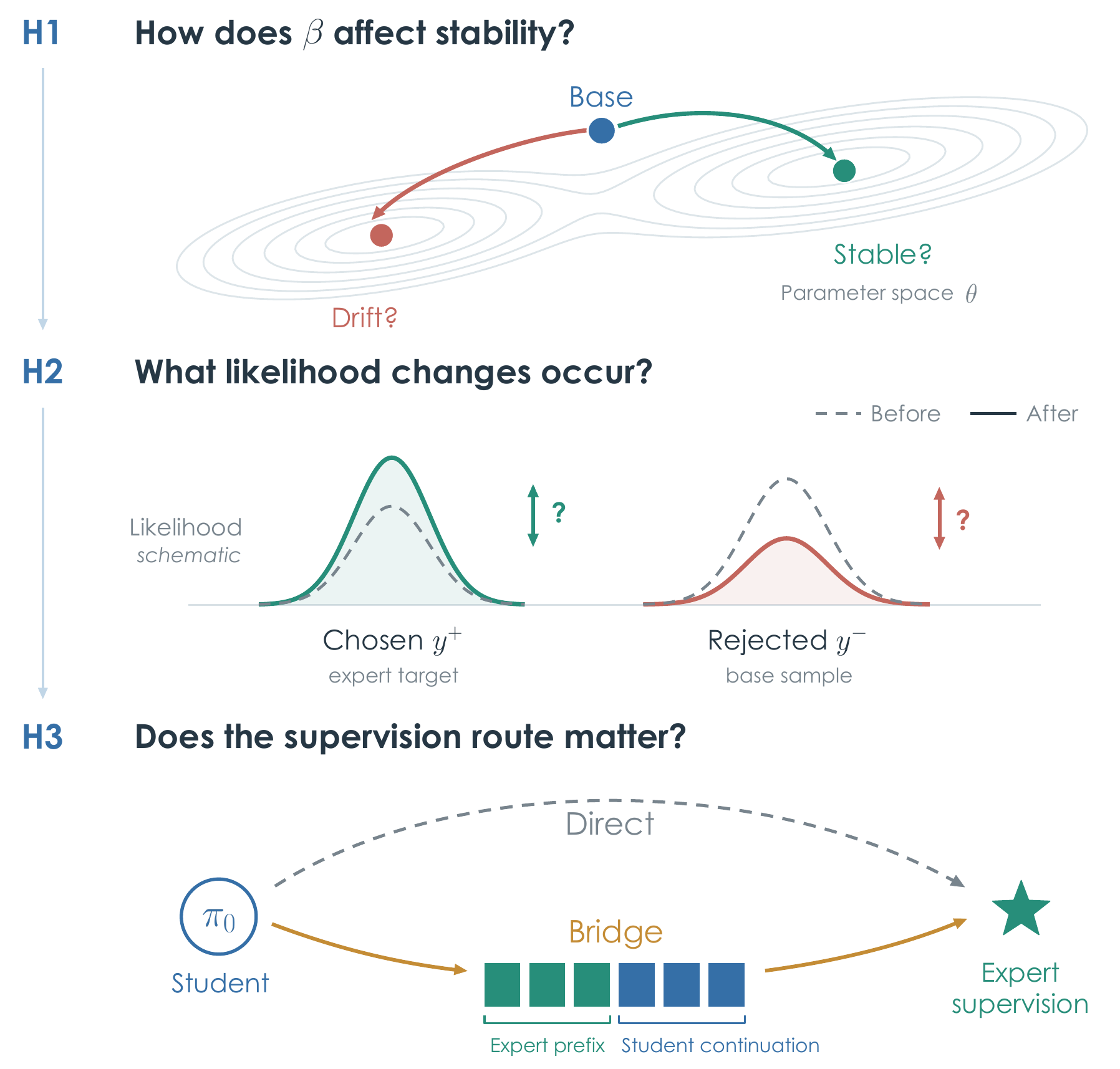}
\caption{Motivating questions for expert-guided preference optimization. Schematic panels H1--H3 correspond to RQ1--RQ3: objective sensitivity, likelihood responses, and supervision routes.}
\label{fig:motivation}
\end{figure}

One option is to rank student generations using recognition or speaker metrics \citep{tian2025preference,hussain2025koel}. Another is to pair student generations with external demonstrations from real recordings or an expert synthesizer. We call the latter \emph{expert-guided preference optimization}. Demonstrations supply preferred targets without ranking a pool of student candidates and can provide examples absent from that pool. The attraction is straightforward: teach the student to favor an expert's response over its own. Whether this produces a better speech generator is less straightforward.

Existing evidence cautions against equating demonstration quality with training benefit. TTS studies report limited gains from ground-truth chosen responses compared with ranked model outputs \citep{tian2025preference,hussain2025koel}, with related observations in non-verbal vocalization synthesis \citep{li2026preference}. Expert speech must also pass through the student's speech-token representation, which can alter its quality. During optimization, a larger preference margin can coexist with declining chosen-response likelihood \citep{pal2024smaug}; such a decline need not imply worse generation \citep{im2026normalized}. These observations leave an empirical question: how does fitting expert targets relate to generation quality under different objectives and supervision routes?

We study CosyVoice and GPT-SoVITS as students, with real recordings and Qwen3-TTS generations as expert sources. We compare DPO, Weighted Preference Optimization (WPO), Anchored Preference Optimization's zero variant (APO-zero), and Noise Contrastive Alignment (NCA) \citep{zhou2024wpo,doosterlinck2025apo,chen2024nca}. RQ1 examines sensitivity to the score scale $\beta$, and RQ2 relates likelihood changes to generation outcomes. RQ3 asks how supervision routes affect these outcomes across objectives, expert sources, and students.

We compare Direct training on full expert targets with a staged \emph{Bridge} route. A Bridge target combines an expert prefix with a student-generated continuation, adapting a construction explored in InCo-DPO \citep{wang2025incodpo} to speech. This route introduces Bridge targets before full expert supervision and refreshes rejected responses between phases. The comparison evaluates the complete protocol. Fixed expert and initial-student probes make likelihood changes comparable across checkpoints even when training pairs change.

Our study yields three findings about expert supervision:
\begin{itemize}
    \item Reconstruction can erode an expert's intelligibility advantage, so source quality alone does not characterize the supervision received by the student.
    \item Training benefits depend on the objective and supervision route: objectives differ in sensitivity to $\beta$, while bridge supervision can mitigate degradation without consistently surpassing the pretrained base.
    \item Expert-probe likelihood can increase while intelligibility worsens, or decrease while it improves; target fitting alone therefore cannot establish generation benefit.
\end{itemize}

%% file: Sections/related.tex
\subsection{Post-Training for Speech Generation}

Speech post-training draws on reinforcement learning from human feedback \citep{ouyang2022instructgpt} and PPO/GRPO policy optimization \citep{schulman2017ppo,shao2024deepseekmath}. Seed-TTS, IndexTTS~2.5, and GLM-TTS optimize rewards for recognition accuracy, speaker consistency, or expressiveness \citep{anastassiou2024seedtts,li2026indextts25,cui2025glmtts}. Other work studies GRPO-based TTS and ASR-guided updates \citep{liu2026grpotts,hussain2026align2speak}, differentiable rewards \citep{gao2025diffro,gao2025explorerl}, and metric optimization for diffusion synthesis and duration prediction \citep{li2025dmospeech,li2026dmospeech2}. At inference, MSpoofTTS uses spoof detectors for pruning and reranking; TED-TTS controls emotion and duration without fine-tuning \citep{zhao2026mspooftts,liang2026tedtts}.

DPO uses preference pairs; KTO uses individual desirability labels \citep{rafailov2023dpo,ethayarajh2024kto}. Speech preferences often rank candidates by recognition or speaker metrics \citep{tian2025preference,hussain2025koel,zhang2025intp}. UNO and RIO study unpaired feedback and reverse-inference supervision \citep{chen2024uno,hu2024rio}; FPO and TKTO study fine-grained and targeted token-level optimization \citep{yao2026fpo,kotoge2026tkto}. Align-SLM uses semantic feedback for speech continuation \citep{lin2025alignslm}, while non-verbal synthesis requires preferences sensitive to lexical content and vocal events \citep{li2026preference}. We compare how objectives use external expert demonstrations.

\subsection{Preference Data and Policy Compatibility}

Discrete latent learning \citep{oord2017vqvae} and neural audio codecs \citep{zeghidour2021soundstream,defossez2022encodec} support token-based audio modeling and reconstruction. AudioLM combines semantic and acoustic tokens for structure and acoustic detail \citep{borsos2022audiolm}; we evaluate expert speech before and after reconstruction through the student's representation.

SpeechAlign pairs encoded recordings with synthesized responses and refreshes the latter \citep{zhang2024speechalign}. However, ground-truth chosen responses yield limited benefits in some TTS studies \citep{tian2025preference,hussain2025koel}; ground-truth and hybrid targets can also underperform synthetic pairs in non-verbal synthesis \citep{li2026preference}. These findings motivate assessing training value separately from source quality.

On-policy sampling and online preference learning connect data--policy compatibility to alignment outcomes \citep{tajwar2024onpolicy,calandriello2024online,tang2024onlineoffline}. WPO reweights offline pairs using current-policy probabilities \citep{zhou2024wpo}. InCo-DPO combines stronger-model prefixes with student continuations and selects pairs using reward-model scores \citep{wang2025incodpo}. We adapt its continuation construction to speech as Bridge. Varying the expert-prefix ratio lets us compare target quality, likelihood, and generation outcomes.

\subsection{Likelihood Dynamics and Generation Quality}

Preference objectives modify likelihoods differently. SLiC-HF and RRHF calibrate or rank sequence likelihoods \citep{zhao2023slichf,yuan2023rrhf}, while IPO provides an alternative preference-learning formulation \citep{azar2024ipo}. ORPO and CPO combine preference learning with likelihood-based fitting, and SimPO uses length-normalized, reference-free rewards \citep{hong2024orpo,xu2024cpo,meng2024simpo}. APO addresses underspecified likelihood updates, and NCA distinguishes absolute likelihood optimization from relative comparisons \citep{doosterlinck2025apo,chen2024nca}. A larger preference margin can coexist with declining chosen likelihood \citep{pal2024smaug,razin2025unalignment}, yet this decline need not imply worse generation \citep{im2026normalized}. Length effects and reward overoptimization further motivate separating training signals from outcome quality \citep{park2024length,gao2023overoptimization}.

NISQA and UTMOS predict perceptual ratings; SpeechBERTScore compares generated and reference representations \citep{mittag2021nisqa,saeki2022utmos,saeki2024speechbertscore}. SpeechLMScore uses discrete-unit likelihood under a separately trained speech language model \citep{maiti2023speechlmscore}. Our diagnostics track the student's likelihood of fixed expert and initial-student probes across checkpoints, relating these changes to generation without equating them with perceptual improvement.

%% file: Sections/study_design.tex
\subsection{Research Questions and Comparison Axes}
\label{subsec:research-questions}

Here, $\beta$ scales the reference-relative preference score, following the standard DPO formulation (Appendix~\ref{sec:app-common-notation}, Eq.~\ref{eq:app_preference_score}). Three research questions examine expert supervision across objectives and routes:
\begin{itemize}
    \item \textbf{RQ1:} How sensitive are different preference objectives to the preference-score scale $\beta$?
    \item \textbf{RQ2:} How do expert and initial-student likelihoods change, and how do these changes relate to generation outcomes?
    \item \textbf{RQ3:} How do Direct--Bridge differences vary across objectives, expert sources, and students?
\end{itemize}
DPO optimizes a relative preference margin, and WPO reweights that comparison \citep{rafailov2023dpo,zhou2024wpo}. APO-zero and NCA act on chosen and rejected scores separately \citep{doosterlinck2025apo,chen2024nca}. We therefore measure both likelihood responses alongside generation outcomes.

\subsection{Characterizing Expert Supervision}
\label{subsec:expert-supervision}

Expert speech enters training through the student's speech-token representation. Table~\ref{tab:expert-supervision} compares original and reconstructed demonstrations with CosyVoice outputs. Qwen3-TTS retains its WER advantage after reconstruction; real recordings largely lose theirs. Rankings also differ across metrics: original recordings have lower WER but lower predicted quality than student generations.

\input{Tables/cosyvoice_ratio}

Source-level preferences guarantee neither better utterances on every metric nor training gains. We assess target quality, initial-policy likelihood, and post-training generation separately.

\subsection{Direct and Bridge Supervision}
\label{subsec:bridge-construction}

Let $x_i$ contain the text and speaker prompt, and $\mathbf y_i^{\mathrm E}$ the expert's student-native tokens. Denote the initial student by $\pi_0$ and the policy entering phase $t$ by $\pi_t$. With $\pi_t$ frozen, sample $\mathbf y^{\mathrm P}_{t,i}\sim\pi_t(\cdot\mid x_i)$. Bold symbols denote sequences; plain symbols denote tokens.

Direct training uses the full expert sequence as the chosen response. Bridge adapts the expert-prefix/student-continuation construction of InCo-DPO to speech \citep{wang2025incodpo}. For an expert-prefix ratio $r$, define $k_i(r)=\lfloor r|\mathbf y_i^{\mathrm E}|\rfloor$. For $0<r<1$,
\begin{equation}
\mathbf y^{\mathrm B}_{t,i}(r)
=\mathbf y^{\mathrm E}_{i,1:k_i(r)}\oplus\widetilde{\mathbf y}_{t,i}(r),
\label{eq:sd-bridge}
\end{equation}
where $\widetilde{\mathbf y}_{t,i}(r)\sim\pi_t(\cdot\mid x_i,\mathbf y^{\mathrm E}_{i,1:k_i(r)})$ and $\oplus$ denotes concatenation. We set $\mathbf y^{\mathrm B}_{t,i}(0)=\mathbf y^{\mathrm P}_{t,i}$ and $\mathbf y^{\mathrm B}_{t,i}(1)=\mathbf y_i^{\mathrm E}$. The ratio measures the fraction of expert tokens retained, not their share of final audio duration.

\begin{figure*}[t]
\centering
\includegraphics[width=0.9\textwidth]{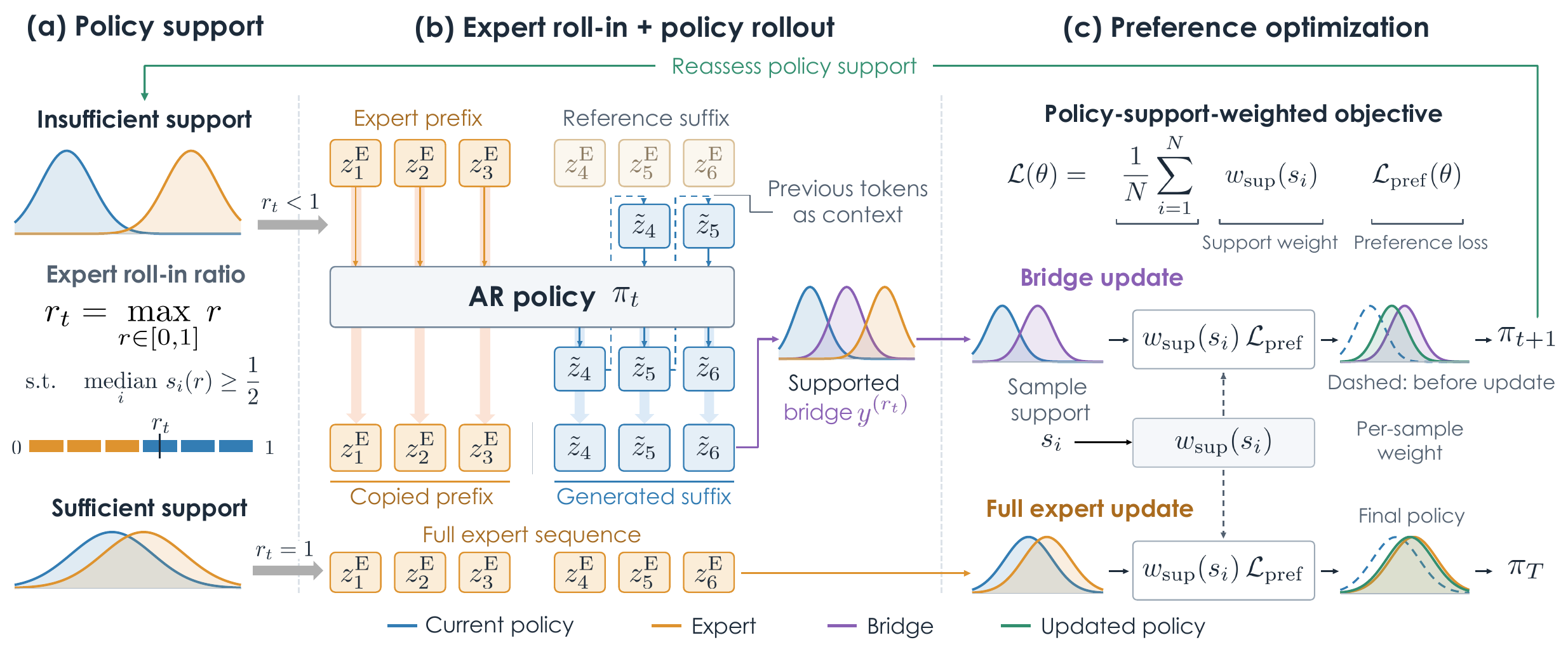}
\caption{Bridge construction and support-weighted training (schematic). Token symbols $z$ correspond to $y$ in the text; experiments use a discrete ratio grid and two fixed training phases.}
\label{fig:bridge-overview}
\end{figure*}

At training ratio $r_t$, Bridge pairs chosen $\mathbf y^+_{t,i}=\mathbf y^{\mathrm B}_{t,i}(r_t)$ with rejected $\mathbf y^-_{t,i}=\mathbf y^{\mathrm P}_{t,i}$. It then uses full expert targets with refreshed student negatives; Direct retains full expert targets and initial-student negatives. Both start from the same student, so the comparison evaluates the complete route, including target composition and negative-sample refresh (Figure~\ref{fig:bridge-overview}).

\subsection{Ratio Selection and Policy-Support Weighting}
\label{subsec:bridge-selection}
\label{subsec:support-weighting}

For a nonempty sequence, length-normalized negative log-likelihood (NLL) is
\begin{equation}
\operatorname{NLL}_{\pi}(\mathbf y\mid x)
=-\frac{1}{|\mathbf y|}\sum_{j=1}^{|\mathbf y|}\log\pi(y_j\mid x,\mathbf y_{<j}).
\label{eq:sd-nll}
\end{equation}
Natural logarithms give nats/token. Teacher-forced scoring covers the full sequence, including copied expert prefixes, with a consistent token mask. Under frozen $\pi_t$, define
\begin{equation}
\begin{aligned}
g_{t,i}(r)&=\operatorname{NLL}_{\pi_t}(\mathbf y^{\mathrm B}_{t,i}(r)\mid x_i)\\
&\quad-\operatorname{NLL}_{\pi_t}(\mathbf y^{\mathrm P}_{t,i}\mid x_i),\\
s_{t,i}(r)&=\min\{1,\exp[-g_{t,i}(r)]\}.
\end{aligned}
\label{eq:sd-support}
\end{equation}
A positive gap means lower per-token likelihood for the target. Before clipping, $\exp(-g)$ is the ratio of geometric-mean token probabilities. The relative-likelihood score $s$, also called the support score, measures neither perceptual quality nor distributional distance.

For the common set $I_t$ of selection inputs with finite scores at every candidate ratio, the selection rule is
\begin{equation}
\begin{aligned}
S_t(r)&=\operatorname{median}_{i\in I_t}s_{t,i}(r),\\
r_t^{\mathrm{sel}}&=\max\{r\in\mathcal A:S_t(r)\geq\tau\},
\end{aligned}
\label{eq:sd-selection}
\end{equation}
with $\mathcal A=\{0,0.2,0.4,0.6,0.8,1\}$ and $\tau=0.5$. Selection uses \texttt{dev-clean}; training targets are constructed separately on \texttt{train-clean-100}. Speech-quality metrics do not enter this rule. Since $S_t(0)=1$, the feasible set is nonempty, although $r=0$ gives identical chosen and rejected responses. The threshold is heuristic. The schematic's $r_t$ denotes $r_t^{\mathrm{sel}}$; actual training ratios are reported in Section~\ref{sec:exp-settings}.

\begin{figure*}[t]
\centering
\includegraphics[page=1,width=\textwidth]{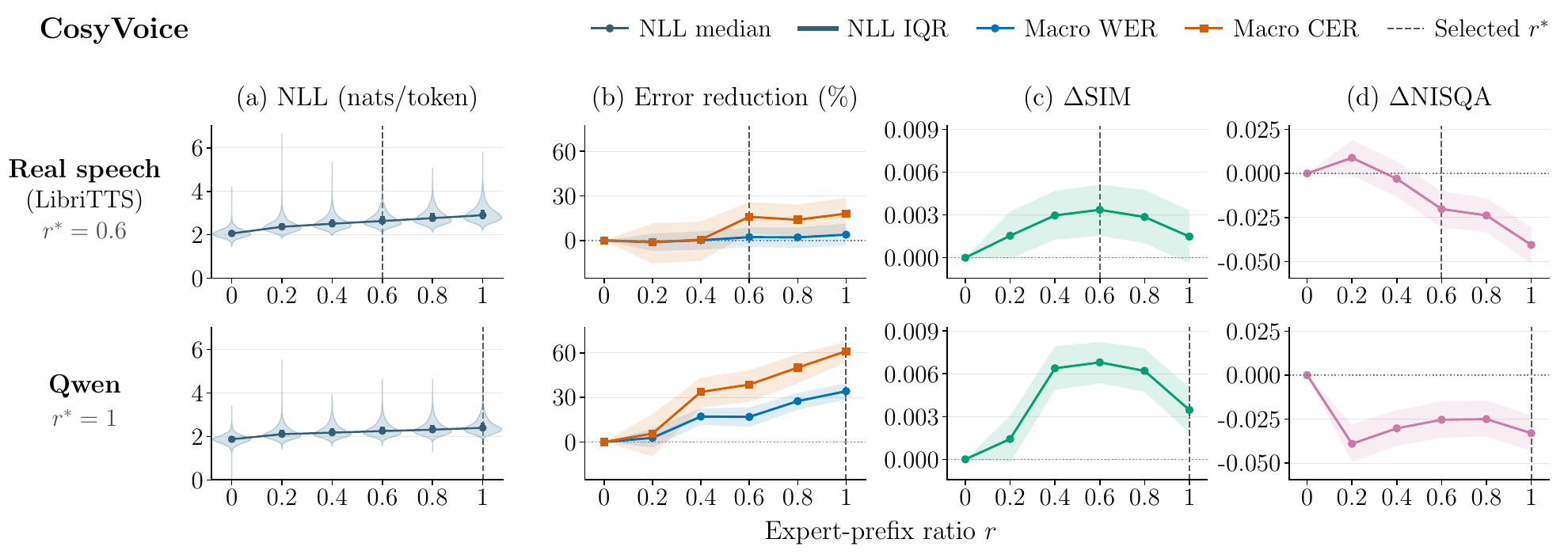}
\caption{CosyVoice constructed targets across ratios. Error reduction uses $r=0$; SIM/NISQA show absolute changes. Dashed lines mark rule-selected ratios $r_t^{\mathrm{sel}}$, distinct from training ratios. Shaded quality bands show 95\% bootstrap intervals.}
\label{fig:bridge-ratio}
\end{figure*}

In Figure~\ref{fig:bridge-ratio}, more expert tokens generally raise target NLL, while quality varies by metric and source. These measurements characterize constructed targets before training.

\paragraph{Policy-support weighting.}
For sample-specific scores $s_i=s_{t,i}(r_t)$ at the training ratio, parabolic weighting defines
\begin{equation}
\begin{aligned}
\mathcal L_t(\theta)&=\frac1N\sum_{i=1}^{N}w_{\mathrm{sup}}(s_i)\mathcal L_{\mathrm{pref},i}(\theta),\\
w_{\mathrm{sup}}(s)&=4s(1-s).
\end{aligned}
\label{eq:support-weighted-objective}
\end{equation}
Scores are computed under $\pi_t$ and fixed during training. The weight peaks at $s=0.5$ and vanishes at $s=0,1$, emphasizing intermediate support. It is separate from the preference objective and WPO's own pair weighting. Clipping maps every $g\leq0$ to $s=1$, so maximal support does not establish that a target is on-policy.

\subsection{Relating Target Fitting to Generation}
\label{subsec:learning-generation}

Within each student--expert setting, we score fixed full-expert and initial-student probes across objectives, routes, and checkpoints. This separates fitting from changes in training targets. For checkpoint $\pi_\theta$, define
\begin{equation}
\begin{aligned}
\delta_\theta(\mathbf y\mid x)
&=\frac{\log\pi_\theta(\mathbf y\mid x)-\log\pi_0(\mathbf y\mid x)}{|\mathbf y|}\\
&=\operatorname{NLL}_{\pi_0}(\mathbf y\mid x)-\operatorname{NLL}_{\pi_\theta}(\mathbf y\mid x).
\end{aligned}
\label{eq:sd-likelihood-change}
\end{equation}
Positive $\delta_\theta$ indicates increased likelihood of the same sequence. The tables report its mean over expert and initial-student probes as $\Delta\ell_{\mathrm E}$ and $\Delta\ell_{\mathrm P}$ (nats/token). The table heading ``Base'' denotes initial-student probes. Both checkpoints score identical conditioning and tokens with the same mask. Unlike the gap $g$, which compares two targets under one policy, $\delta_\theta$ compares two policies on one target. The initial diagnostic baseline $\pi_0$ also remains distinct from the objective's reference policy $\pi_{\mathrm{ref},t}$.

We evaluate intelligibility, speaker similarity, and predicted quality against both Direct and the base, distinguishing mitigation from improvement beyond the starting model. Probe fitting may disagree with generation \citep{tang2024onlineoffline}; likelihood responses alone establish neither causality nor perceptual improvement.

%% file: Tables/cosyvoice_ratio.tex
\begin{table}[t]
\centering
\small
\setlength{\tabcolsep}{4pt}
\begin{tabular}{@{}lrrr@{}}
\toprule
Speech source & WER (\%)$\downarrow$ & SIM$\uparrow$ & NISQA$\uparrow$ \\
\midrule
Pretrained base & 4.2 & 0.709 & 4.357 \\
\addlinespace[2pt]
Real: original & 3.2 & 0.672 & 4.066 \\
Real: reconstructed & 4.1 & 0.721 & 4.349 \\
\addlinespace[2pt]
Qwen: original & 2.7 & 0.753 & 4.358 \\
Qwen: reconstructed & 2.9 & 0.798 & 4.368 \\
\bottomrule
\end{tabular}
\caption{Target quality before CosyVoice training. WER is corpus-level; SIM uses a common speaker reference per input. Reconstruction uses the student's speech representation.}
\label{tab:expert-supervision}
\end{table}

%% file: Sections/experimental_setup.tex
\paragraph{Models and data.}
We use CosyVoice-1 (300M) and GPT-SoVITS v4 as students, with LibriTTS recordings and Qwen3-TTS (1.7B) generations as expert sources \citep{du2024cosyvoice,zen2019libritts,hu2026qwen3tts}. Expert speech is encoded into each student's native representation. Training uses \texttt{train-clean-100}, and ratio assessment uses \texttt{dev-clean}. Generation evaluation uses fixed 4--10-second subsets of LibriTTS \texttt{test-clean} and English SEED TTS Eval \citep{anastassiou2024seedtts}.

\paragraph{Training configurations.}
For RQ1--RQ2, we compare all four objectives at $\beta\in\{0.01,0.1\}$ using CosyVoice with real-speech supervision. For RQ3, we fix $\beta=0.1$ and compare both routes across students and expert sources. Direct trains for two epochs with full expert targets and fixed initial-student negatives. Bridge uses one epoch with partial-expert targets, followed by one epoch with full expert targets and regenerated intermediate-student negatives. First-phase ratios are $r=0.6$ for CosyVoice--real, $r=0.8$ for CosyVoice--Qwen, $r=0.4$ for GPT-SoVITS--real, and $r=0.8$ for GPT-SoVITS--Qwen.

\paragraph{Evaluation.}
We report utterance-level macro WER (\%), speaker similarity (SIM), and NISQA predicted speech quality \citep{mittag2021nisqa}. The ratio plot also reports character error rate (CER). Table~\ref{tab:expert-supervision} separately reports corpus-level WER for target characterization. The two aggregations are not directly interchangeable. Fixed-probe likelihood changes use full-expert and initial-student sequences, independently of phase-specific training pairs. Results are point estimates; we distinguish their magnitude and direction without claiming statistical significance.

Unless accompanied by confidence intervals, results are point estimates and small differences are interpreted cautiously. All NLL measurements, conditional KL measurements, and paired speaker-bootstrap comparisons use the same shared pool of 1,000 utterance IDs across objectives, routes, and checkpoints. Bootstrap intervals quantify uncertainty in generation-metric differences. Conditional KL is measured relative to the original pretrained base on fixed teacher-forced contexts.

%% file: Sections/results.tex
\input{Tables/kl_po}
\input{Tables/cosyvoice_main}
\input{Tables/gpt_sovits_main}

We first examine objective-dependent sensitivity to $\beta$ (RQ1) and
likelihood responses on fixed probes (RQ2), using CosyVoice with
LibriTTS expert supervision. We then compare Direct and Bridge across
both students and expert sources (RQ3), before considering whether
improvements agree across evaluation metrics (RQ2--RQ3). Tables report
WER as a percentage and fixed-probe log-likelihood changes relative
to the initial student; positive changes indicate increased likelihood.
Colors encode this direction, not the quality of generated speech.
Expert rows report reference speech scores; -- marks unreported or inapplicable entries.

\subsection[Objective-Dependent Sensitivity to Beta]{Objective-Dependent Sensitivity to $\beta$}
\label{subsec:results-h1}

Table~\ref{tab:cosyvoice_beta} shows that reducing $\beta$ from $0.1$ to $0.01$ substantially degrades DPO and WPO in this setting.. For DPO, LibriTTS WER rises from $4.59\%$ to $29.78\%$, and SEED TTS Eval WER from $6.33\%$
to $63.02\%$. WPO shows the same pattern, with WER increasing from
$4.81\%$ to $25.57\%$ and from $6.58\%$ to $49.08\%$, respectively.
Both objectives also lose speaker similarity and predicted speech quality.

APO-zero and NCA show smaller changes across the same two values. Their
LibriTTS WER remains between $3.67\%$ and $3.77\%$, compared with
$3.87\%$ for the pretrained base. The tested $\beta$ reduction therefore
has \textbf{markedly different consequences across objectives}. These results
establish local sensitivity in one student--expert setting; two values
do not establish a stability range or an optimal $\beta$ for any objective.
A shared numerical $\beta$ is a comparison condition, not evidence that
different losses induce matched update strengths. We use $\beta=0.1$ in
the remaining comparisons and examine how the contrasting generation
outcomes relate to likelihood changes.

\subsection{Likelihood Responses and Their Limits}
\label{subsec:results-h2}

As shown in Table~\ref{tab:cosyvoice_beta}, at $\beta=0.1$, DPO and WPO decrease mean token log-likelihood on expert probes by $0.348$ and $0.295$, respectively, whereas APO-zero and NCA increase it by $0.083$ and $0.033$. This difference extends beyond the mean:
no evaluated expert probes gain likelihood under DPO or WPO, compared
with $99.8\%$ under APO-zero and $96.0\%$ under NCA. All four objectives,
however, decrease likelihood on $99.9$--$100\%$ of initial-student probes.
Initial-student suppression therefore occurs both when WER improves over
the base and when it deteriorates. Within this setting, expert-probe
responses distinguish these outcomes more clearly than initial-student
suppression.

The expert-probe distinction persists at $\beta=0.01$: the positive-change
rates are $0\%$ for DPO/WPO and $97.7\%/97.6\%$ for APO-zero/NCA.
Yet a large decrease on initial-student probes is not sufficient to
identify poor generation. APO-zero decreases their mean log-likelihood
by $9.243$, more than WPO's $8.446$, while their LibriTTS WERs are
$3.77\%$ and $25.57\%$. \textbf{Neither the prevalence nor the magnitude of
initial-student suppression alone distinguishes these outcomes.}

The expert-probe responses are consistent with the objectives' different
structures. DPO/WPO optimize relative preference scores without requiring
a likelihood increase for the chosen response, while APO-zero and NCA
act on chosen and rejected scores separately.
NCA additionally counterbalances chosen-score growth. This comparison
characterizes the observed updates; it does not establish that the
likelihood changes cause the generation differences. The following
cross-setting comparisons test the limits of interpreting these diagnostics.

\subsection{Route Effects Across Supervision Settings}
\label{subsec:results-h3}

\begin{figure}[t]
\centering
\includegraphics[width=\columnwidth]{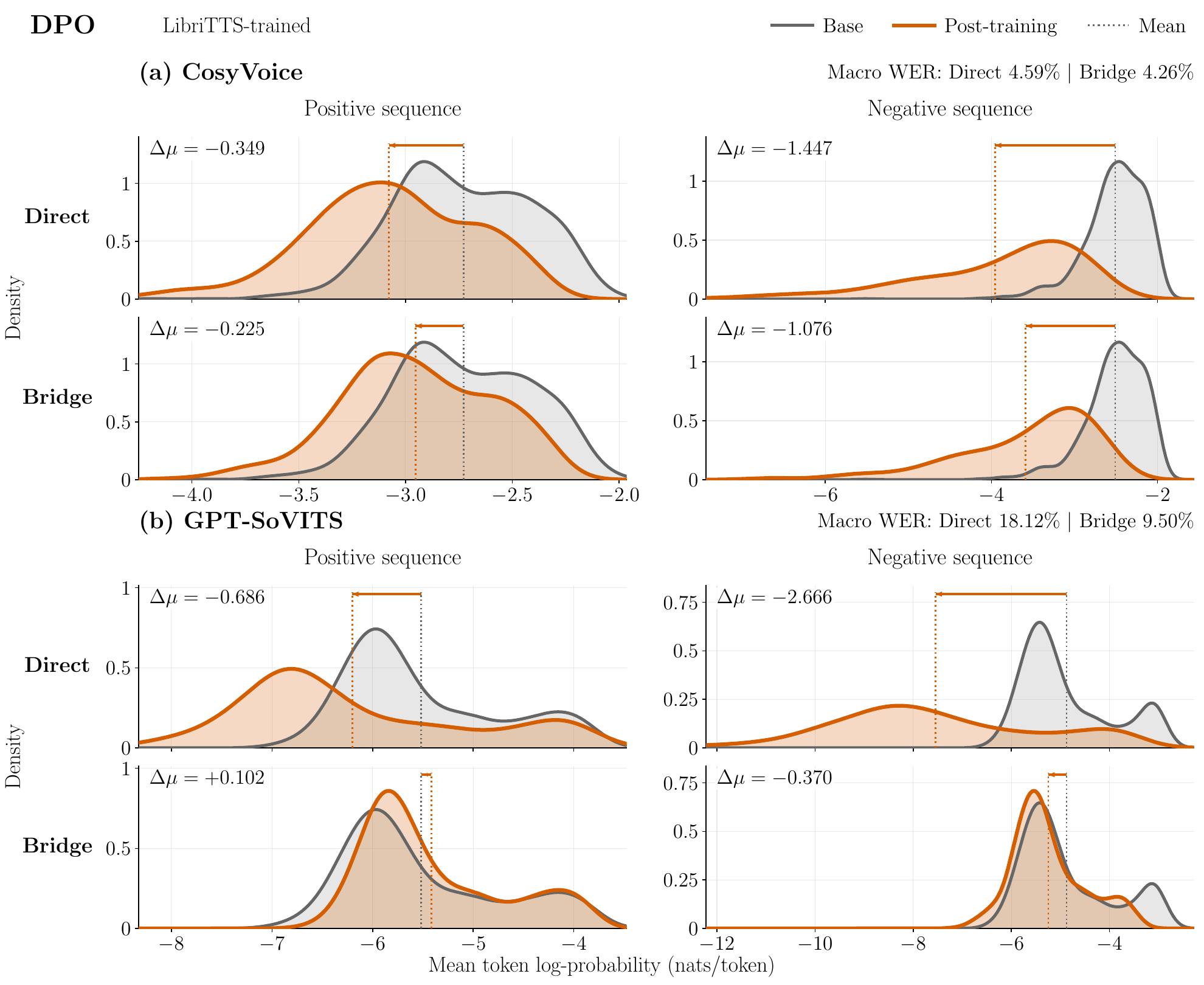}
\caption{DPO log-probability distributions under real-speech supervision. Positive/negative sequences denote fixed expert/initial-student probes. Dotted lines mark means; $\Delta\mu=\mu_{\mathrm{post}}-\mu_{\mathrm{base}}$. Panel headings report macro WER.}
\label{fig:dpo_likelihood_distributions}
\end{figure}

\paragraph{Real-speech supervision.}
Tables~\ref{tab:cosyvoice_results} and~\ref{tab:gpt_sovits_results} show that Bridge reduces the LibriTTS WER deterioration of DPO/WPO for both students.
For CosyVoice, DPO WER decreases from $4.59\%$ under Direct to $4.26\%$
under Bridge, and WPO from $4.81\%$ to $4.18\%$. Both remain above the
base WER of $3.87\%$. For GPT-SoVITS, the corresponding changes are
$18.12\%$ to $9.50\%$ and $13.03\%$ to $9.49\%$, approaching the
base value of $9.50\%$. Figure~\ref{fig:dpo_likelihood_distributions} shows the corresponding DPO distributions. This mitigation accompanies smaller expert-probe
likelihood losses for CosyVoice and a change from loss to gain for
GPT-SoVITS DPO. Yet the latter's expert-probe gain of $0.102$ does not
translate into better-than-base WER: Bridge matches the displayed
LibriTTS baseline and reaches $4.87\%$ on SEED TTS Eval, above the base's
$4.27\%$. \textbf{Recovery from Direct's deterioration should therefore be distinguished from improvement over the starting model.}
Benefits do not extend uniformly across evaluation sets:
GPT-SoVITS WPO worsens on SEED TTS Eval, from $4.58\%$ to $5.49\%$.
APO-zero and NCA likewise show no consistent additional WER benefit from Bridge.

\begin{figure*}[!t]
    \centering
    \includegraphics[width=\textwidth]{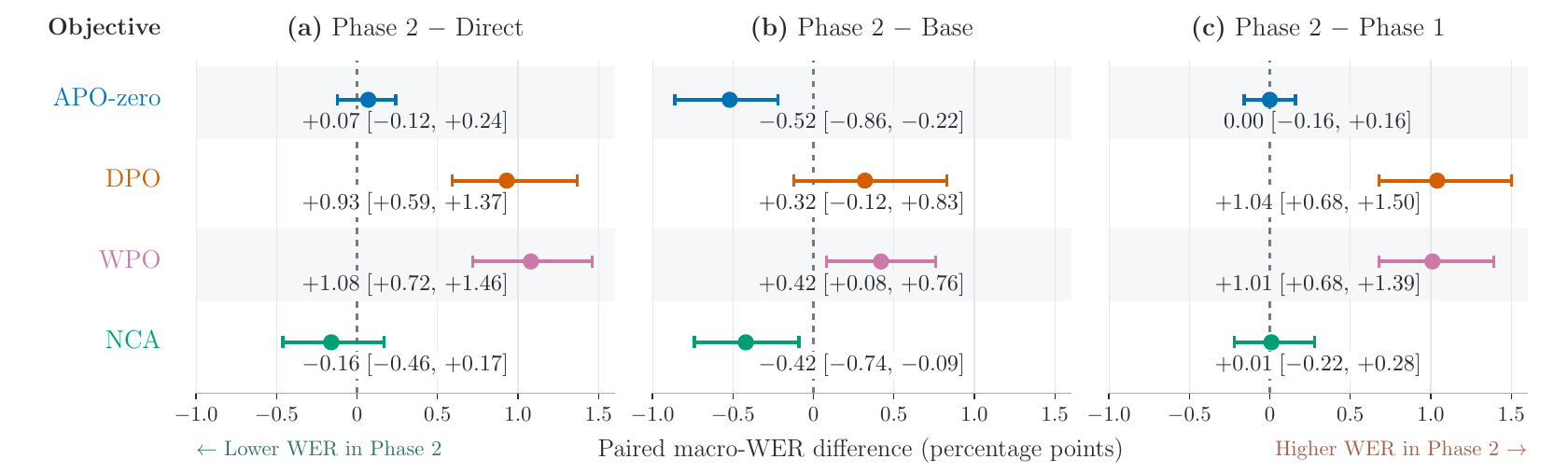}
    \caption{Paired macro-WER differences (Phase~2 minus comparator, in percentage points) with 95\% speaker-bootstrap confidence intervals. Negative values favor Phase~2.}
    \label{fig:bridge_wer_paired}
\end{figure*}

\paragraph{Synthetic-expert supervision.}
With Qwen3-TTS supervision, CosyVoice Direct DPO/WPO improve WER over the
base on both evaluation sets, despite decreasing expert-probe likelihood.
DPO reduces LibriTTS WER from $3.87\%$ to $3.27\%$ and SEED TTS Eval WER
from $4.27\%$ to $2.60\%$, while expert-probe $\Delta\ell=-0.318$.
Bridge reduces this likelihood loss to $-0.164$, but increases WER to
$4.27\%$ and $3.48\%$, respectively. Thus, neither the sign of expert
fitting relative to the base nor its improvement over Direct orders these
generation outcomes. Together with GPT-SoVITS under real-speech supervision,
these cases show why the likelihood--generation association observed in
one setting cannot serve as a general route-selection criterion.

GPT-SoVITS shows a different pattern with the same synthetic expert.
Bridge lowers LibriTTS WER for DPO, WPO, and NCA, but slightly raises it
for APO-zero. SEED TTS Eval changes are also mixed: WPO/NCA improve,
whereas DPO/APO-zero worsen. These comparisons answer RQ3 by identifying
setting-dependent mitigation and gains, rather than a consistently
superior route.

\subsection{Checkpoint Analysis of Bridge Training}
\label{subsec:bridge_checkpoints}

We examine where the CosyVoice--Qwen Bridge route
loses its Phase~1 gains.
Figure~\ref{fig:bridge_wer_paired} reports paired
WER differences on the shared 1,000-utterance pool.
For DPO and WPO, Phase~2 increases WER relative
to Phase~1 by 1.04 percentage points
(95\% CI: $[0.68,1.50]$) and 1.01 points
($[0.68,1.39]$), respectively, and also
underperforms Direct training. Relative to the pretrained base, WPO's WER
increase has an interval excluding zero,
whereas DPO's interval includes zero.
APO-zero and NCA show no clearly resolved phase
difference and retain lower final WER than the base.

The stage-specific likelihood measurements in
Table~\ref{tab:cosyvoice_qwen_checkpoints}
characterize the accompanying updates.
During Phase~2, DPO/WPO substantially decrease
rejected-response likelihood while producing
little net change in mean chosen-target likelihood.
Rejecting generations from the improved intermediate
policy may discourage useful Phase~1 behavior without
a corresponding increase in expert fitting.
However, Phase~1 can improve WER despite declining
chosen likelihood, and NCA remains stable despite
little mean chosen-likelihood change.
Neither rejected-response suppression nor weak
expert fitting alone explains the deterioration.

Table~\ref{tab:bridge_conditional_kl} provides a further counterexample through conditional KL measurements.
Direct DPO/WPO have larger KL to the pretrained
base than their final Bridge checkpoints on both
fixed context sets, yet achieve lower WER.
Thus, conditional KL magnitude alone does not
consistently rank generation quality.


\begin{table}[!t]
    \centering
    \small
    \setlength{\tabcolsep}{3pt}
    \renewcommand{\arraystretch}{1.1}
    \begin{tabular}{@{}lcccccc@{}}
        \toprule
        \multicolumn{1}{c}{\multirow{2}{*}{Objective}}
            & \multicolumn{2}{c}{Phase 1}
            & \multicolumn{2}{c}{Phase 2}
            & \multicolumn{2}{c}{Direct} \\
        \cmidrule(lr){2-3}
        \cmidrule(lr){4-5}
        \cmidrule(l){6-7}
            & $E$ & $S_0$ & $E$ & $S_0$ & $E$ & $S_0$ \\
        \midrule
        APO-zero & 0.088 & 0.165 & 0.113 & 0.331 & 0.120 & 0.299 \\
        DPO      & 0.204 & 0.308 & 0.265 & 0.453 & 0.391 & 0.640 \\
        WPO      & 0.159 & 0.229 & 0.226 & 0.354 & 0.345 & 0.509 \\
        NCA      & 0.019 & 0.058 & 0.016 & 0.098 & 0.018 & 0.099 \\
        \bottomrule
    \end{tabular}
    \caption{
        Conditional KL to the pretrained base (nats/token)
        on fixed expert ($E$) and initial-student ($S_0$) contexts,
        averaged over tokens within each sequence, then over utterances.
        See Appendix~\ref{app:checkpoint_diagnostics}.
    }
    \label{tab:bridge_conditional_kl}
\end{table}

These analyses locate an objective-dependent
deterioration in Phase~2, but do not isolate its
cause: the phase transition jointly changes chosen
targets, refreshes rejected responses, and resets
the optimizer state.

\subsection{Differences Across Evaluation Metrics}
\label{subsec:results-metrics}

Generation metrics also favor different configurations. Under synthetic
supervision, CosyVoice Direct WPO improves LibriTTS WER from the base's
$3.87\%$ to $3.22\%$ and NISQA from $4.484$ to $4.543$, while SIM falls
from $0.731$ to $0.715$. For GPT-SoVITS with real-speech supervision,
Direct APO-zero improves SEED TTS Eval WER from $4.27\%$ to $3.83\%$
and SIM from $0.401$ to $0.411$, but lowers NISQA from $3.986$ to $3.940$.
\textbf{Intelligibility gains therefore need not coincide with gains in speaker similarity or predicted speech quality.} A single claim of ``better
speech'' would conceal these trade-offs.

Across RQ2--RQ3, fixed-probe diagnostics describe target fitting, while
generation metrics establish observed benefits. Their relationship varies
by student, expert source, route, and evaluation set; neither expert
likelihood nor a single generation metric fully ranks the tested configurations.

%% file: Tables/kl_po.tex
\begin{table*}[t]
\centering
\caption{CosyVoice with LibriTTS supervision under Direct training. Green/red marks likelihood increases/decreases. Bold marks the best student result across objective--$\beta$ settings.}
\label{tab:cosyvoice_beta}
\begingroup
\footnotesize
\setlength{\tabcolsep}{3.2pt}
\renewcommand{\arraystretch}{0.96}
\setlength{\aboverulesep}{0.35ex}
\setlength{\belowrulesep}{0.35ex}
\begin{tabular*}{0.96\textwidth}{@{\extracolsep{\fill}}llrrrrrrr@{\hspace{10pt}}r@{}}
\toprule
\multirow{2}{*}[-0.4ex]{Objective} & \multirow{2}{*}[-0.4ex]{$\beta$} & \multicolumn{3}{c}{LibriTTS test} & \multicolumn{3}{c}{SEED TTS Eval} & \multicolumn{2}{c}{Fixed probes: $\Delta\ell$} \\
\cmidrule(lr){3-5}\cmidrule(lr){6-8}\cmidrule(l){9-10}
& & WER(\%)$\downarrow$ & SIM$\uparrow$ & NISQA$\uparrow$ & WER(\%)$\downarrow$ & SIM$\uparrow$ & NISQA$\uparrow$ & Expert & Base \\
\midrule
Pretrained base & -- & 3.87 & 0.731 & 4.484 & 4.27 & 0.739 & 4.169 & $0.000$ & $0.000$ \\
\textit{LibriTTS expert} & -- & 3.40 & 0.696 & 4.151 & -- & -- & -- & -- & -- \\
\midrule
\multirow{2}{*}{DPO} & 0.1 & 4.59 & 0.706 & 4.427 & 6.33 & 0.721 & 4.057 & \llloss{-0.348} & \llloss{-1.448} \\
 & 0.01 & 29.78 & 0.611 & 4.119 & 63.02 & 0.496 & 3.462 & \llloss{-3.473} & \llloss{-22.301} \\
\multirow{2}{*}{WPO} & 0.1 & 4.81 & 0.708 & 4.429 & 6.58 & 0.722 & 4.072 & \llloss{-0.295} & \llloss{-1.191} \\
 & 0.01 & 25.57 & 0.559 & 4.161 & 49.08 & 0.530 & 3.552 & \llloss{-3.467} & \llloss{-8.446} \\
\multirow{2}{*}{APO-zero} & 0.1 & 3.73 & 0.730 & 4.487 & 3.70 & 0.734 & \textbf{4.172} & \llgain{+0.083} & \llloss{-0.516} \\
 & 0.01 & 3.77 & 0.731 & 4.491 & 3.77 & \textbf{0.737} & 4.155 & \llgain{+0.056} & \llloss{-9.243} \\
\multirow{2}{*}{NCA} & 0.1 & 3.76 & \textbf{0.732} & \textbf{4.501} & 3.74 & 0.736 & 4.171 & \llgain{+0.033} & \llloss{-0.420} \\
 & 0.01 & \textbf{3.67} & 0.730 & 4.489 & \textbf{3.39} & 0.736 & 4.143 & \llgain{+0.061} & \llloss{-7.078} \\
\bottomrule
\end{tabular*}
\endgroup
\end{table*}

%% file: Tables/cosyvoice_main.tex
\begin{table*}[t]
\centering
\caption{CosyVoice: Direct versus Bridge at $\beta=0.1$. Green/red marks likelihood increases/decreases. Bold marks the best student result within each expert block.}
\label{tab:cosyvoice_results}
\begingroup
\footnotesize
\setlength{\tabcolsep}{3.2pt}
\renewcommand{\arraystretch}{0.96}
\setlength{\aboverulesep}{0.35ex}
\setlength{\belowrulesep}{0.35ex}
\begin{tabular*}{0.96\textwidth}{@{\extracolsep{\fill}}llrrrrrrr@{\hspace{10pt}}r@{}}
\toprule
\multirow{2}{*}[-0.4ex]{Objective} & \multirow{2}{*}[-0.4ex]{Route} & \multicolumn{3}{c}{LibriTTS test} & \multicolumn{3}{c}{SEED TTS Eval} & \multicolumn{2}{c}{Fixed probes: $\Delta\ell$} \\
\cmidrule(lr){3-5}\cmidrule(lr){6-8}\cmidrule(l){9-10}
& & WER(\%)$\downarrow$ & SIM$\uparrow$ & NISQA$\uparrow$ & WER(\%)$\downarrow$ & SIM$\uparrow$ & NISQA$\uparrow$ & Expert & Base \\
\midrule
Pretrained base & -- & 3.87 & 0.731 & 4.484 & 4.27 & 0.739 & 4.169 & $0.000$ & $0.000$ \\
\textit{LibriTTS} & \textit{Expert} & 3.40 & 0.696 & 4.151 & -- & -- & -- & -- & -- \\
\textit{Qwen3-TTS} & \textit{Expert} & 2.80 & 0.780 & 4.441 & -- & -- & -- & -- & -- \\
\midrule
\multicolumn{10}{@{}l}{\textit{Expert: LibriTTS recordings}} \\
\multirow{2}{*}{DPO} & Direct & 4.59 & 0.706 & 4.427 & 6.33 & 0.721 & 4.057 & \llloss{-0.348} & \llloss{-1.448} \\
 & Bridge & 4.26 & 0.716 & 4.439 & 5.77 & 0.725 & 4.091 & \llloss{-0.225} & \llloss{-1.077} \\
\multirow{2}{*}{WPO} & Direct & 4.81 & 0.708 & 4.429 & 6.58 & 0.722 & 4.072 & \llloss{-0.295} & \llloss{-1.191} \\
 & Bridge & 4.18 & 0.726 & 4.496 & 4.87 & 0.728 & \textbf{4.181} & \llloss{-0.101} & \llloss{-0.773} \\
\multirow{2}{*}{APO-zero} & Direct & \textbf{3.73} & 0.730 & 4.487 & \textbf{3.70} & 0.734 & 4.172 & \llgain{+0.083} & \llloss{-0.516} \\
 & Bridge & 3.80 & \textbf{0.732} & \textbf{4.509} & 3.72 & 0.735 & 4.162 & \llgain{+0.049} & \llloss{-0.500} \\
\multirow{2}{*}{NCA} & Direct & 3.76 & \textbf{0.732} & 4.501 & 3.74 & 0.736 & 4.171 & \llgain{+0.033} & \llloss{-0.420} \\
 & Bridge & 3.78 & 0.731 & 4.508 & 3.97 & \textbf{0.737} & 4.169 & \llgain{+0.022} & \llloss{-0.391} \\
\midrule
\multicolumn{10}{@{}l}{\textit{Expert: Qwen3-TTS synthesis}} \\
\multirow{2}{*}{DPO} & Direct & 3.27 & 0.712 & 4.542 & 2.60 & 0.732 & 4.247 & \llloss{-0.318} & \llloss{-1.400} \\
 & Bridge & 4.27 & 0.723 & 4.498 & 3.48 & 0.727 & 4.121 & \llloss{-0.164} & \llloss{-0.961} \\
\multirow{2}{*}{WPO} & Direct & \textbf{3.22} & 0.715 & 4.543 & 3.29 & 0.730 & \textbf{4.265} & \llloss{-0.270} & \llloss{-1.117} \\
 & Bridge & 4.29 & 0.725 & 4.507 & 3.79 & 0.727 & 4.129 & \llloss{-0.127} & \llloss{-0.774} \\
\multirow{2}{*}{APO-zero} & Direct & 3.32 & 0.723 & \textbf{4.549} & 2.65 & 0.731 & 4.220 & \llgain{+0.152} & \llloss{-0.562} \\
 & Bridge & 3.38 & 0.726 & 4.545 & \textbf{2.55} & 0.734 & 4.202 & \llgain{+0.150} & \llloss{-0.711} \\
\multirow{2}{*}{NCA} & Direct & 3.62 & \textbf{0.731} & 4.520 & 3.24 & \textbf{0.739} & 4.198 & \llgain{+0.038} & \llloss{-0.341} \\
 & Bridge & 3.48 & 0.728 & 4.511 & 3.38 & \textbf{0.739} & 4.180 & \llgain{+0.039} & \llloss{-0.349} \\
\bottomrule
\end{tabular*}
\endgroup
\end{table*}

%% file: Tables/gpt_sovits_main.tex
\begin{table*}[t]
\centering
\caption{GPT-SoVITS: Direct versus Bridge at $\beta=0.1$. Green/red marks likelihood increases/decreases. Bold marks the best student result within each expert block.}
\label{tab:gpt_sovits_results}
\begingroup
\footnotesize
\setlength{\tabcolsep}{3.2pt}
\renewcommand{\arraystretch}{0.96}
\setlength{\aboverulesep}{0.35ex}
\setlength{\belowrulesep}{0.35ex}
\begin{tabular*}{0.96\textwidth}{@{\extracolsep{\fill}}llrrrrrrr@{\hspace{10pt}}r@{}}
\toprule
\multirow{2}{*}[-0.4ex]{Objective} & \multirow{2}{*}[-0.4ex]{Route} & \multicolumn{3}{c}{LibriTTS test} & \multicolumn{3}{c}{SEED TTS Eval} & \multicolumn{2}{c}{Fixed probes: $\Delta\ell$} \\
\cmidrule(lr){3-5}\cmidrule(lr){6-8}\cmidrule(l){9-10}
& & WER(\%)$\downarrow$ & SIM$\uparrow$ & NISQA$\uparrow$ & WER(\%)$\downarrow$ & SIM$\uparrow$ & NISQA$\uparrow$ & Expert & Base \\
\midrule
Pretrained base & -- & 9.50 & 0.387 & 3.767 & 4.27 & 0.401 & 3.986 & $0.000$ & $0.000$ \\
\textit{LibriTTS} & \textit{Expert} & 3.40 & 0.696 & 4.151 & -- & -- & -- & -- & -- \\
\textit{Qwen3-TTS} & \textit{Expert} & 2.80 & 0.780 & 4.441 & -- & -- & -- & -- & -- \\
\midrule
\multicolumn{10}{@{}l}{\textit{Expert: LibriTTS recordings}} \\
\multirow{2}{*}{DPO} & Direct & 18.12 & 0.381 & 3.577 & 5.25 & 0.406 & 3.908 & \llloss{-0.685} & \llloss{-2.666} \\
 & Bridge & 9.50 & 0.387 & \textbf{3.767} & 4.87 & 0.400 & 3.931 & \llgain{+0.102} & \llloss{-0.370} \\
\multirow{2}{*}{WPO} & Direct & 13.03 & 0.381 & 3.563 & 4.58 & 0.405 & 3.912 & \llloss{-0.030} & \llloss{-0.504} \\
 & Bridge & \textbf{9.49} & 0.387 & \textbf{3.767} & 5.49 & 0.398 & 3.916 & $0.000$ & \llloss{-0.429} \\
\multirow{2}{*}{APO-zero} & Direct & 11.78 & \textbf{0.392} & 3.696 & \textbf{3.83} & \textbf{0.411} & 3.940 & \llgain{+0.138} & \llloss{-0.414} \\
 & Bridge & 9.50 & 0.387 & \textbf{3.767} & 4.98 & 0.388 & 3.903 & \llgain{+0.081} & \llloss{-0.860} \\
\multirow{2}{*}{NCA} & Direct & 9.71 & 0.390 & 3.754 & 3.97 & 0.406 & \textbf{4.003} & \llgain{+0.018} & \llloss{-0.235} \\
 & Bridge & 9.50 & 0.387 & \textbf{3.767} & 4.65 & 0.398 & 3.969 & \llgain{+0.005} & \llloss{-0.417} \\
\midrule
\multicolumn{10}{@{}l}{\textit{Expert: Qwen3-TTS synthesis}} \\
\multirow{2}{*}{DPO} & Direct & 7.54 & 0.483 & 3.823 & 4.46 & 0.406 & 3.931 & \llloss{-0.033} & \llloss{-0.539} \\
 & Bridge & 7.01 & 0.477 & 3.854 & 4.64 & 0.400 & 3.976 & \llloss{-0.034} & \llloss{-0.495} \\
\multirow{2}{*}{WPO} & Direct & 7.59 & 0.478 & 3.812 & 4.77 & 0.403 & 3.943 & \llloss{-0.027} & \llloss{-0.520} \\
 & Bridge & 7.46 & 0.485 & 3.862 & 4.35 & 0.405 & 3.977 & \llgain{+0.028} & \llloss{-0.245} \\
\multirow{2}{*}{APO-zero} & Direct & \textbf{6.73} & \textbf{0.493} & 3.900 & \textbf{4.00} & \textbf{0.410} & 3.946 & \llgain{+0.235} & \llloss{-0.287} \\
 & Bridge & 6.76 & 0.489 & 3.890 & 4.70 & 0.408 & 3.969 & \llgain{+0.152} & \llloss{-0.182} \\
\multirow{2}{*}{NCA} & Direct & 7.20 & 0.486 & 3.900 & 4.52 & 0.401 & \textbf{3.979} & \llgain{+0.014} & \llloss{-0.208} \\
 & Bridge & 7.08 & 0.486 & \textbf{3.906} & 4.14 & 0.404 & 3.977 & \llgain{+0.011} & \llloss{-0.090} \\
\bottomrule
\end{tabular*}
\endgroup
\end{table*}

%% file: Sections/conclusion.tex
Expert supervision for TTS requires more than selecting an appealing demonstration. Our study distinguishes the quality of expert speech, the student's fitting of fixed targets, and the quality of its subsequent generations. Across the evaluated settings, these quantities do not provide interchangeable evidence of improvement. Preference objectives respond differently to the tested score scales, and staged Bridge supervision can mitigate degradation without surpassing the pretrained base. Most directly, increased expert likelihood can accompany worse intelligibility, while decreased likelihood can accompany better generation. The practical implication is to assess expert supervision through complementary generation metrics and explicit comparisons with the starting model. Fixed-probe likelihoods help describe what optimization changes; their value lies in interpreting those changes alongside generated speech, rather than certifying training success.

%% file: Sections/limitations.tex
Our study examines two student models and two expert sources on English utterances. The findings are scoped to these models, sources, and evaluation settings; broader language and duration coverage remains for future work. Generation results are point estimates from task-specific comparisons, so small differences warrant cautious interpretation. Our route comparison uses a fixed two-phase Bridge protocol; more stages and adaptive supervision schedules remain to be explored.

%% file: Appendix/appendix.tex
\section{Background of Preference Optimization Algorithms}
\label{sec:background}
\input{Appendix/Sections/background}

\section{Training Datasets Statistics}
\label{sec:bridges}
\input{Appendix/Sections/bridges}

\section{Implementation Details}
\label{sec:implementation}
\input{Appendix/Sections/implementation}

\section{Evaluation Metrics}
\label{sec:metrics}
\input{Appendix/Sections/metrics}

\section{Supplementary Results}
\label{sec:suplementary}
\input{Appendix/Sections/supplementary}

\section{AI Usage Statement}
Large language models were used to assist with language refinement, manuscript organization, and LaTeX formatting. All AI-assisted content was reviewed by the authors for accuracy and consistency with the study’s methods and experimental findings. The authors take full responsibility for the accuracy and integrity of the final manuscript.

%% file: Appendix/Sections/background.tex
We summarize the four preference optimization objectives
considered in this study: DPO, WPO, APO-zero, and NCA.

\subsection{Common Notation}
\label{sec:app-common-notation}

Let $(x,\mathbf{y}^{+},\mathbf{y}^{-})$ denote a preference
triple, where $x$ is the conditioning input and
$\mathbf{y}^{+}$ and $\mathbf{y}^{-}$ are the chosen and
rejected speech-token sequences, respectively.
Let $\pi_\theta$ be the trainable policy and
$\pi_{\mathrm{ref}}$ a frozen reference policy.
The sequence log-likelihood is
\begin{equation}
\log \pi_\theta(\mathbf{y}\mid x)
=
\sum_{j=1}^{|\mathbf{y}|}
\log \pi_\theta(y_j\mid x,\mathbf{y}_{<j}).
\end{equation}
Following the standard reference-based formulation, define
\begin{equation}
u_\theta(x,\mathbf{y})
=
\beta
\log
\frac{\pi_\theta(\mathbf{y}\mid x)}
     {\pi_{\mathrm{ref}}(\mathbf{y}\mid x)},
\label{eq:app_preference_score}
\end{equation}
where $\beta>0$ scales the reference-relative score.
We write $u^{+}=u_\theta(x,\mathbf{y}^{+})$,
$u^{-}=u_\theta(x,\mathbf{y}^{-})$, and
$\sigma(z)=(1+e^{-z})^{-1}$.
Each training objective averages the corresponding
per-pair loss over the preference dataset.

\subsection{Direct Preference Optimization (DPO)}

DPO~\citep{rafailov2023dpo} derives a direct policy-learning
objective from KL-regularized reward maximization and a
Bradley--Terry preference model. Its per-pair loss is
\begin{equation}
\mathcal{L}_{\mathrm{DPO}}
=
-\log \sigma(u^{+}-u^{-}).
\label{eq:app_dpo}
\end{equation}
This formulation allows preference learning without training
a separate reward model or performing an explicit
reinforcement-learning rollout during each update.

Intuitively, DPO encourages the chosen response to improve
relative to the rejected response, after accounting for their
likelihoods under the reference policy. Its objective depends
on the score difference rather than either score individually.
Consequently, a larger preference margin can arise from
increasing chosen likelihood, decreasing rejected likelihood,
or decreasing both while suppressing the rejected response
more strongly. DPO therefore does not explicitly require
chosen likelihood to increase relative to the reference.

\subsection{Weighted Preference Optimization (WPO)}

WPO~\citep{zhou2024wpo} addresses the discrepancy between
offline preference data and the current policy by assigning
policy-dependent weights to preference pairs. Its DPO-based
form is
\begin{equation}
\mathcal{L}_{\mathrm{WPO}}
=
\operatorname{sg}[a_\theta]
\mathcal{L}_{\mathrm{DPO}},
\label{eq:app_wpo}
\end{equation}
where $a_\theta$ is a nonnegative pair weight and
$\operatorname{sg}$ denotes stop-gradient.
A basic length-normalized weighting construction is
\begin{equation}
\begin{split}
q_\theta(x,\mathbf{y})
&=
\exp\left[
\frac{1}{|\mathbf{y}|}
\log\pi_\theta(\mathbf{y}\mid x)
\right],
\\
a_\theta
&=
q_\theta(x,\mathbf{y}^{+})
q_\theta(x,\mathbf{y}^{-}).
\end{split}
\end{equation}
The original method additionally introduces calibrated weights
that account for variation in policy confidence.

The intuition is to give greater influence to pairs that
are more compatible with the current policy, approximating
aspects of on-policy learning while retaining an offline
dataset. Detaching the weights prevents the model from
reducing the loss by directly manipulating the weighting
factor. WPO changes the relative contribution of training
pairs but retains DPO's relative-margin objective.

\subsection{Anchored Preference Optimization (APO-zero)}

APO~\citep{doosterlinck2025apo} introduces objectives
that specify the intended directions of chosen and rejected
likelihood changes. We consider its APO-zero variant:
\begin{equation}
\mathcal{L}_{\mathrm{APO\text{-}zero}}
=
1-\sigma(u^{+})+\sigma(u^{-}).
\label{eq:app_apo_zero}
\end{equation}
The additive constant does not affect optimization.

Unlike DPO, APO-zero acts on the two scores separately.
It encourages chosen likelihood to increase relative to the
reference and rejected likelihood to decrease. The zero
anchor represents unchanged likelihood under the reference
policy. Its sigmoid terms yield bounded losses and gradients
that diminish as the individual scores move away from the
anchor. This variant is motivated by settings in which the
chosen responses are desirable examples for the policy
to learn from.

\subsection{Noise Contrastive Alignment (NCA)}

NCA~\citep{chen2024nca} formulates policy alignment through
noise-contrastive estimation. Whereas DPO compares responses
through a relative score difference, NCA learns individual
reference-relative scores. Its binary-preference form is
\begin{equation}
\begin{split}
\mathcal{L}_{\mathrm{NCA}}
={}&
-\log\sigma(u^{+})
\\
&-\frac{1}{2}\log\sigma(-u^{+})
-\frac{1}{2}\log\sigma(-u^{-}).
\end{split}
\label{eq:app_nca}
\end{equation}
The first term supports the chosen response, while the
remaining terms account for the noise contribution of both
responses. The chosen-response noise term counterbalances
unrestricted score growth.

Viewed as a loss over independent scores, NCA has a finite
chosen-score optimum at $u^{+}=\log 2$, while encouraging
the rejected score to decrease. It therefore provides
chosen support with a counterbalancing term, rather than
optimizing only the pairwise margin.

\subsection{Connection to Expert Supervision and Training Stability}

The objectives above differ in whether they optimize
a relative margin, reweight that comparison, or act
on individual reference-relative scores. These
differences motivate examining chosen and rejected
likelihood changes separately. However, objective-level
incentives do not guarantee particular likelihood
changes in a shared neural policy or improvements
in generated-speech quality.

Existing TTS studies likewise question whether
higher-quality demonstrations necessarily provide
better preference supervision.
\citet{tian2025preference} find that ranked
model-generated pairs outperform ground-truth chosen
responses, and report deterioration after an additional
round of iterative optimization. Koel-TTS reports
degeneration when ground-truth speech supplies chosen
responses under its default DPO configuration
\citep{hussain2025koel}. These observations show that
neither ground-truth targets nor additional optimization
rounds guarantee better generation.

Our study extends this evidence across objectives,
students, expert sources, and Direct versus Bridge
routes. We jointly examine demonstration quality after
reconstruction, likelihood changes on fixed probes,
and generated-speech quality. The comparisons identify
settings in which expert fitting and intelligibility
move in opposite directions, and distinguish mitigation
of degradation from improvement over the pretrained
base. Checkpoint analysis further localizes the loss
of Bridge's initial gains and examines whether
conditional KL consistently ranks generation quality.

%% file: Appendix/Sections/bridges.tex
\input{Appendix/Tables/data}
\begin{figure}[!t]
    \centering
    \includegraphics[width=\columnwidth]
        {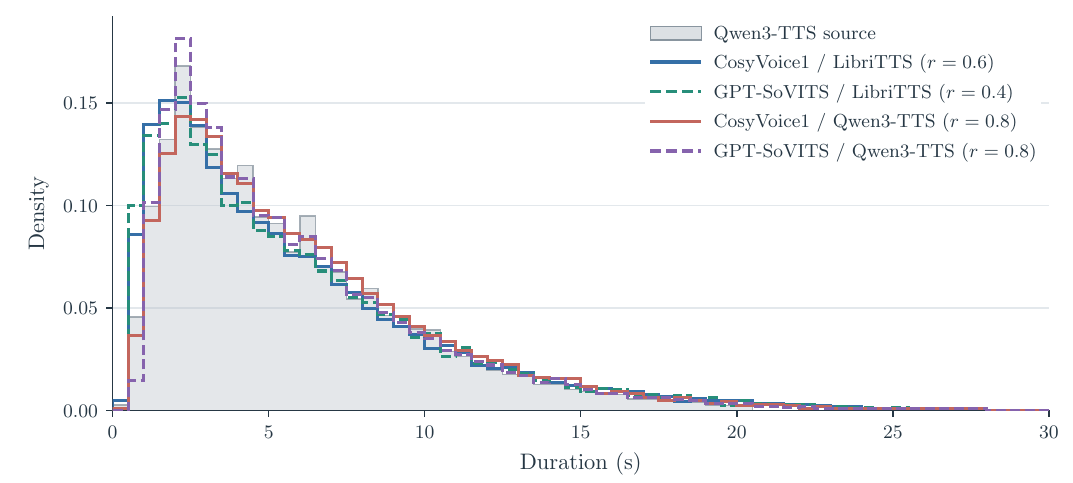}
    \caption{
        Normalized duration histograms of the retained
        Bridge samples across student--expert configurations.
        The Qwen3-TTS source distribution is shown in gray.
        Ratios denote the copied fraction of expert tokens.
    }
    \label{fig:bridge_duration_distributions}
\end{figure}

For each training input, we encode the expert speech into
the student's native speech-token representation.
We retain the first $r$ fraction of expert tokens and
generate the continuation using the incoming student policy,
conditioned on the same text, speaker prompt, and copied prefix.
The concatenated sequence forms the Bridge target, which
is decoded into speech for duration and quality evaluation. 
To maintain a comparable training-set size across
configurations, we retain 36,000 constructed
utterances per Bridge dataset and around 57--58 hours of speech.
Table~\ref{tab:bridge_statistics} reports their composition.
Figure~\ref{fig:bridge_duration_distributions} shows
right-skewed duration distributions, with broadly overlapping
densities across configurations and the Qwen3-TTS source.

\begin{figure*}[!t]
    \centering
    \includegraphics[width=\textwidth]
        {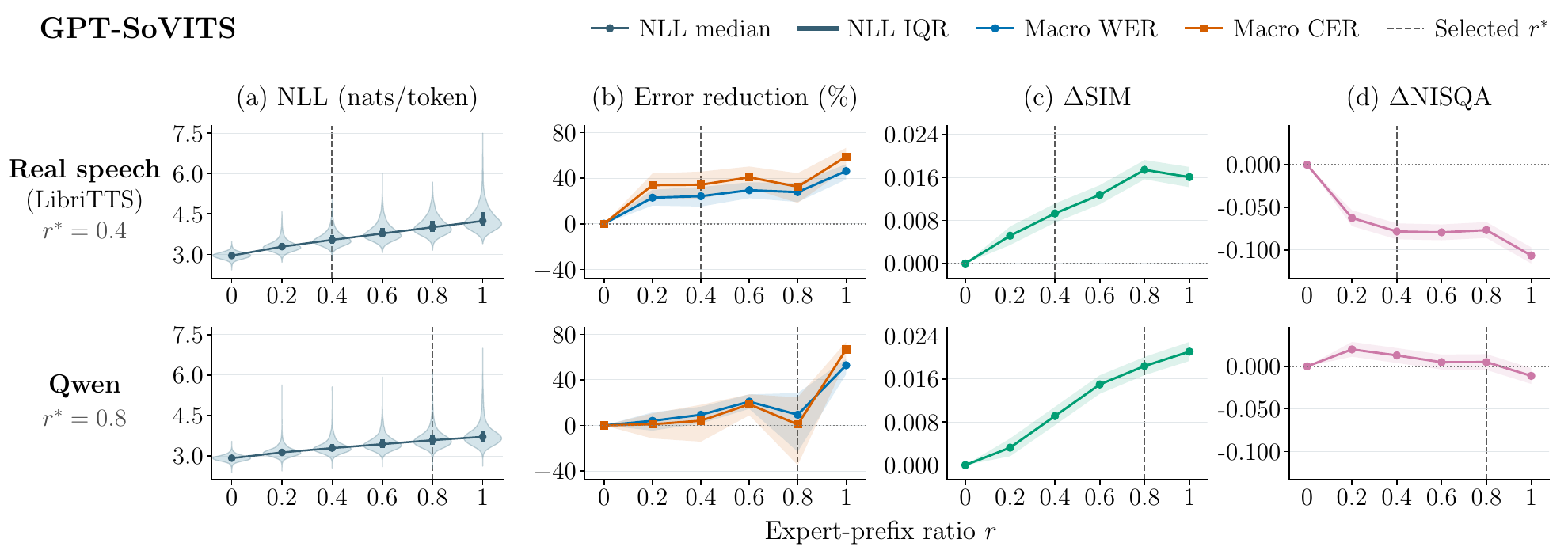}
    \caption{GPT-SoVITS constructed targets across ratios. Error reduction uses $r=0$; SIM/NISQA show absolute changes. Dashed lines mark rule-selected ratios $r_t^{\mathrm{sel}}$, distinct from training ratios. Shaded quality bands show 95\% bootstrap intervals.}
    \label{fig:gpt_sovits_ratio_analysis}
\end{figure*}

Complementing the CosyVoice-1 analysis in the main text,
Figure~\ref{fig:gpt_sovits_ratio_analysis} presents
the corresponding expert-prefix ratio analysis for GPT-SoVITS.
We examine token-level negative log-likelihood (NLL),
reductions in word and character error rates,
and changes in speaker similarity and perceptual quality.
Together, these metrics characterize the trade-offs associated
with varying the proportion of copied expert tokens.

%% file: Appendix/Tables/data.tex
\begin{table*}[!t]
    \centering
    \small
    \renewcommand{\arraystretch}{1.2}
    \caption{
        Composition of the retained Bridge datasets.
        The copied token fraction $r$ specifies the proportion
        of expert tokens retained as the prefix.
        Total speech duration is partitioned into copied expert
        prefixes and student-generated continuations.
    }
    \label{tab:bridge_statistics}
    \begin{tabular}{@{}cccccccc@{}}
        \toprule
        \multirow{2}{*}{Student}
        & \multirow{2}{*}{Expert}
        & \multirow{2}{*}{$r$}
        & \multirow{2}{*}{Utterances}
        & \multirow{2}{*}{Speakers}
        & \multicolumn{3}{c}{Duration (h)} \\
        \cmidrule(lr){6-8}
        & & & & & Total & Prefix & Continuation \\
        \midrule
        CosyVoice-1 & LibriTTS & 0.6
            & 36,000 & 274 & 57.40 & 34.79 & 22.61 \\
        GPT-SoVITS & LibriTTS & 0.4
            & 35,729 & 270 & 57.91 & 22.81 & 35.10 \\
        CosyVoice-1 & Qwen3-TTS & 0.8
            & 33,931 & 267 & 57.05 & 44.72 & 12.33 \\
        GPT-SoVITS & Qwen3-TTS & 0.8
            & 35,459 & 267 & 57.79 & 45.72 & 12.07 \\
        \bottomrule
    \end{tabular}
\end{table*}

%% file: Appendix/Sections/implementation.tex
\paragraph{Model configurations.}
We initialize the students from the publicly released
CosyVoice-1 300M\footnote{
\url{https://huggingface.co/FunAudioLLM/CosyVoice-300M}}
and GPT-SoVITS v4\footnote{
\url{https://huggingface.co/lj1995/GPT-SoVITS}}
checkpoints.
Preference training performs full parameter fine-tuning
of the autoregressive speech-token generation model:
the LLM in CosyVoice and the GPT/Text2Semantic model
in GPT-SoVITS.
The remaining speech-processing and waveform-generation
components are kept fixed.
Synthetic expert speech is generated using
Qwen3-TTS-12Hz-1.7B-Base.\footnote{
\url{https://huggingface.co/Qwen/Qwen3-TTS-12Hz-1.7B-Base}}

\paragraph{Optimization settings.}
Both students use Adam with a learning rate of $10^{-6}$,
batch size 2, zero weight decay, no gradient accumulation,
and no warm-up.
Gradients are clipped to a maximum norm of 5.
Direct training runs for two epochs, while Bridge training
uses one epoch per phase.
We use $\beta=0.1$ unless explicitly varied in the
$\beta$ ablation.
The objectives operate on summed sequence log-probabilities,
including speech tokens, copied expert prefixes, and EOS,
while excluding conditioning positions and padding.
Parabolic policy-support weighting is applied across
objectives and training routes, except when explicitly
varied in the weighting ablation.

\paragraph{Training routes.}
All objectives and routes use the same retained training
inputs within each student--expert setting.
Direct pairs full expert targets with initial-student
generations.
Bridge first trains on constructed Bridge targets paired
with initial-student generations, then trains on full
expert targets paired with negatives regenerated by the
corresponding intermediate student.
Phase 2 starts from the Phase 1 model parameters with
a newly initialized optimizer.
The reference policy remains a frozen copy of the original
pretrained student throughout both phases.

%% file: Appendix/Sections/metrics.tex
We follow CosyVoice 3's CV3-Eval evaluation tools for
content consistency and speaker similarity\footnote{
\url{https://github.com/FunAudioLLM/CV3-Eval}}.
All metrics are averaged across evaluated utterances.

\paragraph{Word Error Rate (WER).}
We transcribe generated English speech using
Whisper-large-v3\footnote{
\url{https://huggingface.co/openai/whisper-large-v3}}
and compare it with the intended text.
WER is the number of word substitutions, deletions,
and insertions divided by the reference word count.
We report macro WER, the average of utterance-level
error rates. Lower is better.

\paragraph{Speaker Similarity (SIM).}
We compute cosine similarity between speaker embeddings
of generated and reference speech using ERes2Net\footnote{
\url{https://github.com/modelscope/3D-Speaker}}.
The speaker reference for each input is fixed across
configurations. Higher is better.

\paragraph{Speech Quality (NISQA).}
We use NISQA~\citep{mittag2021nisqa}, a non-intrusive
speech-quality predictor, to estimate overall perceptual
quality without a reference recording.
Higher scores indicate better predicted quality.

%% file: Appendix/Sections/supplementary.tex


\subsection{Fixed-Probe Likelihood Diagnostics}
\label{sec:full-diagnostics}
Tables~\ref{tab:diagnostics_beta}--\ref{tab:diagnostics_gpt_sovits}
characterize likelihood changes on fixed full-expert and
initial-student probes. Within each student--expert setting,
all checkpoints score the same sequences against a common
initial-student baseline. This keeps the comparison consistent
even when the training pairs change between phases.
Lower NLL indicates higher likelihood.
Expert$\uparrow$ reports the percentage of expert probes
gaining likelihood, while Base$\downarrow$ reports the
percentage of initial-student probes losing likelihood.
Mean NLL and direction rates provide complementary information:
the former captures the aggregate fitting response, while
the latter shows how widely a direction of change is shared
across probes. Green and red indicate likelihood increases
and decreases, respectively. These diagnostics describe
expert fitting and initial-student suppression; their
implications for training benefit must be assessed alongside
generation outcomes.
\input{Appendix/Tables/diagnostics_beta}
\input{Appendix/Tables/diagnostics_cosyvoice}
\input{Appendix/Tables/diagnostics_gpt_sovits}

\subsection{Policy-Support Weighting Ablation}
\label{app:support_weighting}
\input{Appendix/Tables/weights}

We compare constant, linear, and parabolic policy-support
weighting with preference targets and training budgets held
fixed as shown in Table~\ref{tab:support_weighting}.
The observed differences are small, and constant weighting
matches or outperforms both non-uniform variants across the
reported metrics.
Thus, this ablation does not demonstrate a benefit from
non-uniform weighting in the evaluated setting.
We retain parabolic weighting, 
as the default configuration used elsewhere in the paper.
Here, $s$ is the relative-likelihood score defined in the
main text. This weighting emphasizes intermediate scores
and downweights both extremes; it is a design choice,
not an empirically superior variant established by this ablation.

\subsection{Likelihood Distributions Across Preference Objectives}
\label{app:likelihood_distributions}

\begin{figure}[!t]
\centering
\includegraphics[width=\columnwidth]{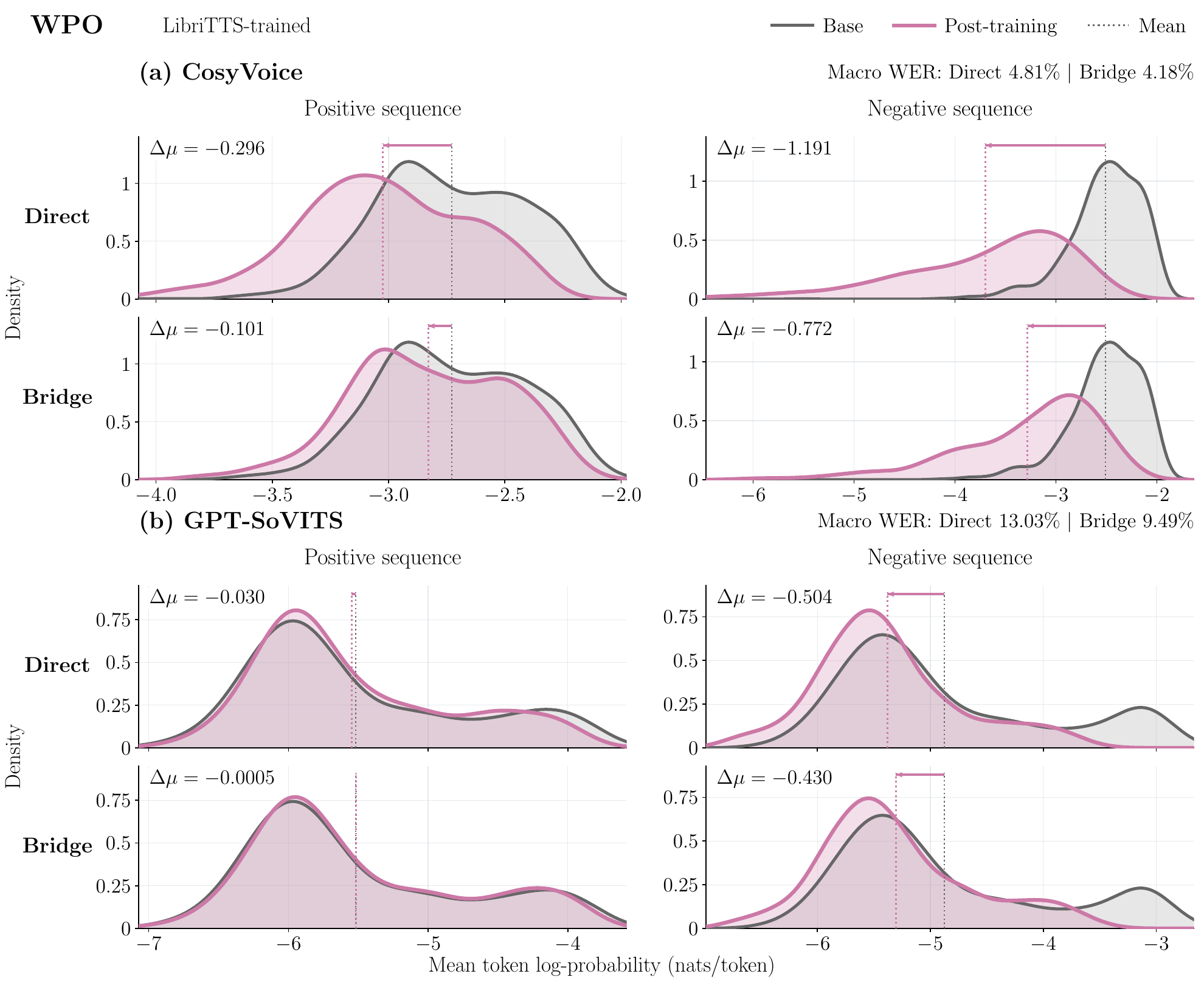}
\caption{WPO fixed-probe log-probability distributions under real-speech supervision.}
\label{fig:wpo_likelihood_distributions}
\end{figure}

\begin{figure}[!t]
\centering
\includegraphics[width=\columnwidth]{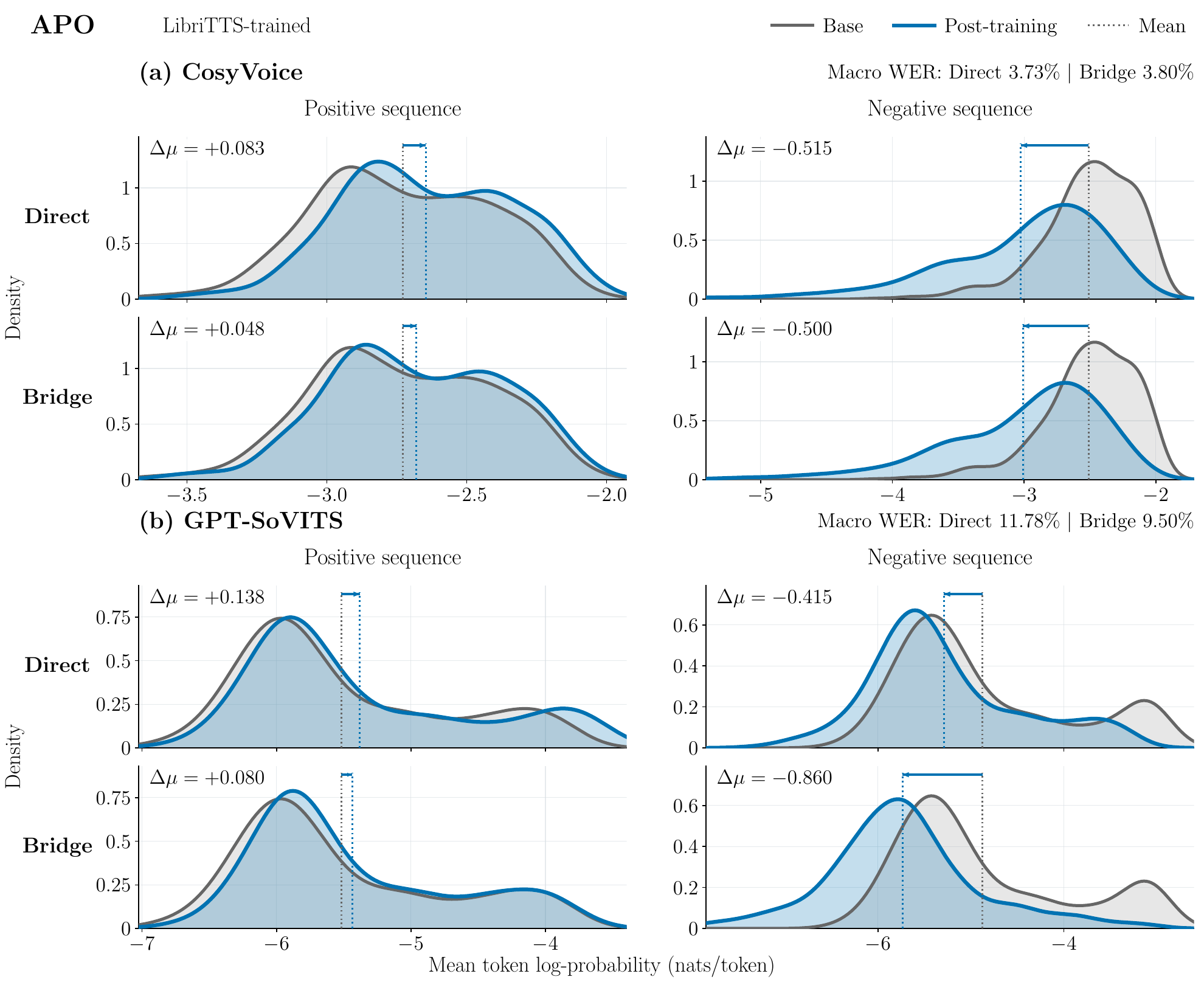}
\caption{APO-zero fixed-probe log-probability distributions under real-speech supervision.}
\label{fig:apo_likelihood_distributions}
\end{figure}

\begin{figure}[!t]
\centering
\includegraphics[width=\columnwidth]{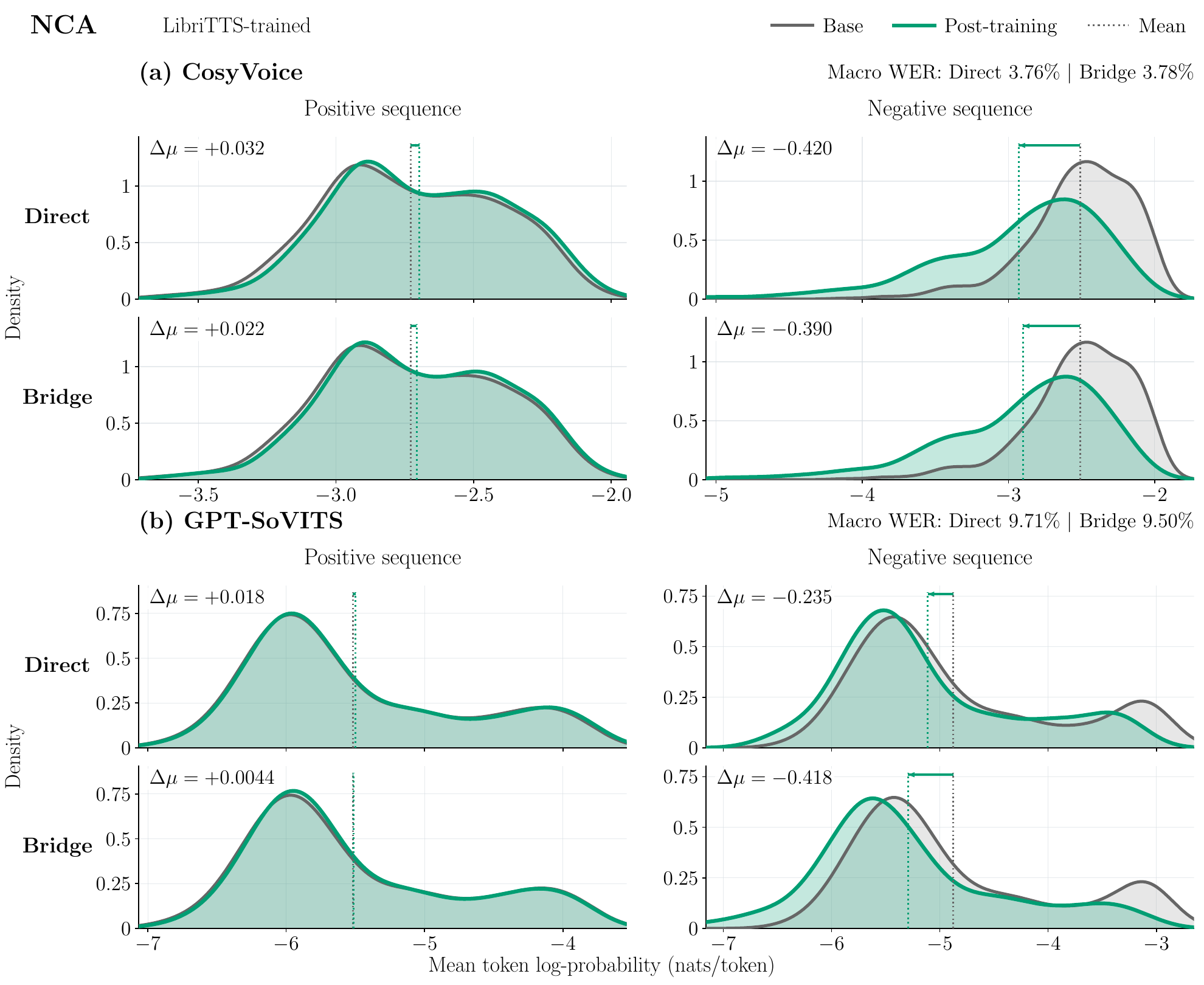}
\caption{NCA fixed-probe log-probability distributions under real-speech supervision.}
\label{fig:nca_likelihood_distributions}
\end{figure}

Figures~\ref{fig:wpo_likelihood_distributions}--\ref{fig:nca_likelihood_distributions}, together with Figure~\ref{fig:dpo_likelihood_distributions}, compare all four objectives under LibriTTS real-speech supervision at $\beta=0.1$. Each curve summarizes mean token log-probabilities on fixed sequences: ``positive'' denotes full-expert probes and ``negative'' denotes initial-student probes. Gray curves show the pretrained base, colored curves the trained policy, and dotted lines their means. The plotted $\Delta\mu$ measures the mean change from the base. Panel headings report LibriTTS macro WER; ``APO'' denotes APO-zero.
For CosyVoice, DPO and WPO lower mean log-probability on both probe sets, with smaller decreases under Bridge. APO-zero and NCA instead raise the expert mean while lowering the initial-student mean. Their expert mean gains are smaller under Bridge than under Direct, indicating weaker expert fitting. Thus, shared initial-student suppression can accompany either increased or decreased expert likelihood.

The route effects differ for GPT-SoVITS. Bridge reverses DPO's expert mean shift from negative to positive, whereas WPO's expert mean shifts remain small under both routes, with a near-zero Bridge change. APO-zero and NCA retain positive expert mean shifts, but Bridge reduces these gains while strengthening initial-student suppression. Bridge therefore does not uniformly reduce the magnitude of likelihood changes across objectives and probe types.
These distributional differences do not provide a universal ranking of generation outcomes. For GPT-SoVITS APO-zero, Bridge lowers LibriTTS WER despite weaker expert fitting and stronger initial-student suppression than Direct, but raises WER on SEED TTS Eval (Table~\ref{tab:gpt_sovits_results}). For CosyVoice APO-zero and NCA, weaker expert fitting under Bridge accompanies slightly higher LibriTTS WER (Table~\ref{tab:cosyvoice_results}). Fixed-probe distributions characterize how each objective and route changes target fitting; generation benefits still require evaluation against both Direct and the pretrained base.

\input{Appendix/Tables/phase_analysis}

\subsection{Intermediate Checkpoint Analysis}
\label{app:intermediate_checkpoints}

Table~\ref{tab:cosyvoice_qwen_checkpoints} examines intermediate
and final checkpoints of staged Bridge training for CosyVoice
under Qwen3-TTS supervision.
Phase~1 reduces macro WER from the pretrained base's $3.91\%$
to $3.19$--$3.48\%$ across the four objectives.
However, subsequent training on full expert targets increases
WER from $3.19\%$ to $4.23\%$ for DPO and from $3.31\%$ to
$4.33\%$ for WPO, leaving both final checkpoints with higher
WER than the base and their Direct counterparts.
APO-zero and NCA remain comparatively stable, with WER
changing by only $0.01$ percentage points across phases.

The Phase~2 likelihood statistics distinguish how the
objectives respond to full expert targets.
DPO and WPO slightly increase mean chosen NLL
($2.423\rightarrow2.427$ and $2.378\rightarrow2.390$),
with reported rejected-suppression rates of $100.0\%$.
Nevertheless, chosen likelihood increases on $47.4\%$ and
$43.1\%$ of examples, respectively, showing that the small
mean shifts summarize mixed update directions.
NCA also produces little change in mean chosen NLL,
with gains on $50.7\%$ of examples.
APO-zero exhibits a more consistent increase in chosen
likelihood: mean NLL decreases from $2.158$ to $2.113$,
with gains on $98.8\%$ of examples.
These checkpoint comparisons locate the observed DPO/WPO
WER deterioration in Phase~2, without establishing its cause.
Because preference pairs and reference checkpoints differ
across stages, the likelihood statistics characterize updates
within each stage. They should not be interpreted as a
fixed-trajectory comparison across checkpoints.

\subsection{Paired Uncertainty and Conditional KL}
\label{app:checkpoint_diagnostics}

All analyses use the shared pool of 1,000 utterances.

\paragraph{Paired WER comparisons.}
We compute percentile 95\% intervals from 10,000
paired speaker-bootstrap replicates.
Each replicate samples speakers with replacement,
retains their utterances with multiplicity, and
recomputes macro-WER differences using identical
resampled IDs across systems.
Intervals capture evaluation-sample uncertainty,
not training-seed variability.

Figure~\ref{fig:bridge_wer_paired}  confirms that
DPO/WPO lose their Phase~1 gains and underperform
Direct training.
Their Phase~2 WER increases relative to Phase~1
are 1.04 percentage points
(95\% CI: $[0.68,1.50]$) and 1.01 points
($[0.68,1.39]$), respectively.
Relative to the pretrained base, WPO's interval
excludes zero, whereas DPO's includes it.
APO-zero/NCA retain lower WER than the base,
but their intervals establish neither a Phase~2
benefit nor equivalence between phases.


\paragraph{Conditional distribution drift.}
Each checkpoint scores identical conditioning
inputs and fixed expert/initial-student sequences.
Let $p_{\theta,i,j}$ and $p_{0,i,j}$ denote its
next-token distribution and the pretrained base's
distribution at position $j$ of utterance $i$.
For each context set, we compute
\begin{equation}
    \widehat{D}_{\theta}
    =
    \frac{1}{N}
    \sum_{i=1}^{N}
    \frac{1}{T_i}
    \sum_{j=1}^{T_i}
    D_{\mathrm{KL}}
    \left(
        p_{\theta,i,j}
        \,\Vert\,
        p_{0,i,j}
    \right).
\end{equation}
Here, $N=1{,}000$ and $T_i$ counts scored positions.
We use pre-sampling decoder distributions over
speech tokens and EOS, excluding conditioning
positions and padding.
Natural logarithms give nats/token.
These measurements describe conditional KL on
fixed contexts, not sequence-policy KL;
differences in KL-to-base are not pairwise
checkpoint KL.

Table~\ref{tab:bridge_conditional_kl} shows that
increased KL to the base is not specific to
Phase~2 deterioration.
APO-zero's initial-student-context KL rises from
0.165 to 0.331 without a clearly resolved WER change.
NCA's expert-context KL decreases from 0.019 to
0.016 while its initial-student-context KL increases
from 0.058 to 0.098.
Together with Direct DPO/WPO's larger KL but lower
WER, these observations show that drift depends
on the scored contexts and does not consistently
order observed generation quality.

%% file: Appendix/Tables/diagnostics_beta.tex
\begin{table}[!t]
\centering
\caption{CosyVoice: fixed-probe diagnostics across $\beta$ under LibriTTS supervision.}
\label{tab:diagnostics_beta}
\begingroup
\footnotesize
\setlength{\tabcolsep}{2pt}
\renewcommand{\arraystretch}{0.98}
\begin{tabular*}{\columnwidth}{@{\extracolsep{\fill}}llrrrr@{}}
\toprule
\multirow{2}{*}[-0.5ex]{Objective} & \multirow{2}{*}[-0.5ex]{$\beta$} & \multicolumn{2}{c}{NLL} & \multicolumn{2}{c}{Probes (\%)} \\
\cmidrule(lr){3-4}\cmidrule(l){5-6}
& & Expert & Base & Expert$\uparrow$ & Base$\downarrow$ \\
\midrule
\multicolumn{2}{@{}l}{Pretrained base} & 2.729 & 2.509 & -- & -- \\
\cmidrule(lr){1-6}
\multirow{2}{*}{DPO} & 0.1 & \llloss{3.077} & \llloss{3.957} & 0.0 & 100.0 \\
 & 0.01 & \llloss{6.202} & \llloss{24.810} & 0.0 & 100.0 \\
\multirow{2}{*}{WPO} & 0.1 & \llloss{3.024} & \llloss{3.700} & 0.0 & 100.0 \\
 & 0.01 & \llloss{6.196} & \llloss{10.955} & 0.0 & 100.0 \\
\multirow{2}{*}{APO-zero} & 0.1 & \llgain{2.646} & \llloss{3.025} & 99.8 & 99.9 \\
 & 0.01 & \llgain{2.673} & \llloss{11.752} & 97.7 & 100.0 \\
\multirow{2}{*}{NCA} & 0.1 & \llgain{2.696} & \llloss{2.929} & 96.0 & 100.0 \\
 & 0.01 & \llgain{2.668} & \llloss{9.587} & 97.6 & 100.0 \\
\bottomrule
\end{tabular*}
\endgroup
\end{table}

%% file: Appendix/Tables/diagnostics_cosyvoice.tex
\begin{table}[!t]
\centering
\caption{CosyVoice: fixed-probe diagnostics by route ($\beta=0.1$).}
\label{tab:diagnostics_cosyvoice}
\begingroup
\footnotesize
\setlength{\tabcolsep}{2pt}
\renewcommand{\arraystretch}{0.98}
\begin{tabular*}{\columnwidth}{@{\extracolsep{\fill}}llrrrr@{}}
\toprule
\multirow{2}{*}[-0.5ex]{Objective} & \multirow{2}{*}[-0.5ex]{Route} & \multicolumn{2}{c}{NLL} & \multicolumn{2}{c}{Probes (\%)} \\
\cmidrule(lr){3-4}\cmidrule(l){5-6}
& & Expert & Base & Expert$\uparrow$ & Base$\downarrow$ \\
\midrule
\multicolumn{6}{@{}l}{\textit{Expert: LibriTTS recordings}} \\[1pt]
\multicolumn{2}{@{}l}{Pretrained base} & 2.729 & 2.509 & -- & -- \\
\cmidrule(lr){1-6}
\multirow{2}{*}{DPO} & Direct & \llloss{3.077} & \llloss{3.957} & 0.0 & 100.0 \\
 & Bridge & \llloss{2.954} & \llloss{3.586} & 0.1 & 100.0 \\
\multirow{2}{*}{WPO} & Direct & \llloss{3.024} & \llloss{3.700} & 0.0 & 100.0 \\
 & Bridge & \llloss{2.830} & \llloss{3.282} & 1.7 & 100.0 \\
\multirow{2}{*}{APO-zero} & Direct & \llgain{2.646} & \llloss{3.025} & 99.8 & 99.9 \\
 & Bridge & \llgain{2.680} & \llloss{3.009} & 97.7 & 99.7 \\
\multirow{2}{*}{NCA} & Direct & \llgain{2.696} & \llloss{2.929} & 96.0 & 100.0 \\
 & Bridge & \llgain{2.707} & \llloss{2.900} & 94.2 & 100.0 \\
\midrule
\multicolumn{6}{@{}l}{\textit{Expert: Qwen3-TTS synthesis}} \\[1pt]
\multicolumn{2}{@{}l}{Pretrained base} & 2.263 & 2.527 & -- & -- \\
\cmidrule(lr){1-6}
\multirow{2}{*}{DPO} & Direct & \llloss{2.581} & \llloss{3.927} & 0.8 & 100.0 \\
 & Bridge & \llloss{2.427} & \llloss{3.488} & 3.5 & 100.0 \\
\multirow{2}{*}{WPO} & Direct & \llloss{2.533} & \llloss{3.644} & 1.2 & 100.0 \\
 & Bridge & \llloss{2.390} & \llloss{3.301} & 6.3 & 100.0 \\
\multirow{2}{*}{APO-zero} & Direct & \llgain{2.111} & \llloss{3.089} & 99.9 & 100.0 \\
 & Bridge & \llgain{2.113} & \llloss{3.238} & 99.9 & 100.0 \\
\multirow{2}{*}{NCA} & Direct & \llgain{2.225} & \llloss{2.868} & 95.5 & 100.0 \\
 & Bridge & \llgain{2.224} & \llloss{2.876} & 98.6 & 100.0 \\
\bottomrule
\end{tabular*}
\endgroup
\end{table}

%% file: Appendix/Tables/diagnostics_gpt_sovits.tex
\begin{table}[!t]
\centering
\caption{GPT-SoVITS: fixed-probe diagnostics by route ($\beta=0.1$).}
\label{tab:diagnostics_gpt_sovits}
\begingroup
\footnotesize
\setlength{\tabcolsep}{2pt}
\renewcommand{\arraystretch}{0.98}
\begin{tabular*}{\columnwidth}{@{\extracolsep{\fill}}llrrrr@{}}
\toprule
\multirow{2}{*}[-0.5ex]{Objective} & \multirow{2}{*}[-0.5ex]{Route} & \multicolumn{2}{c}{NLL} & \multicolumn{2}{c}{Probes (\%)} \\
\cmidrule(lr){3-4}\cmidrule(l){5-6}
& & Expert & Base & Expert$\uparrow$ & Base$\downarrow$ \\
\midrule
\multicolumn{6}{@{}l}{\textit{Expert: LibriTTS recordings}} \\[1pt]
\multicolumn{2}{@{}l}{Pretrained base} & 5.518 & 4.876 & -- & -- \\
\cmidrule(lr){1-6}
\multirow{2}{*}{DPO} & Direct & \llloss{6.203} & \llloss{7.542} & 6.2 & 100.0 \\
 & Bridge & \llgain{5.416} & \llloss{5.246} & 84.8 & 93.5 \\
\multirow{2}{*}{WPO} & Direct & \llloss{5.548} & \llloss{5.380} & 49.8 & 99.2 \\
 & Bridge & 5.518 & \llloss{5.305} & 56.5 & 99.6 \\
\multirow{2}{*}{APO-zero} & Direct & \llgain{5.380} & \llloss{5.290} & 98.2 & 98.9 \\
 & Bridge & \llgain{5.437} & \llloss{5.736} & 87.2 & 99.1 \\
\multirow{2}{*}{NCA} & Direct & \llgain{5.500} & \llloss{5.111} & 74.7 & 99.4 \\
 & Bridge & \llgain{5.513} & \llloss{5.293} & 58.3 & 99.4 \\
\midrule
\multicolumn{6}{@{}l}{\textit{Expert: Qwen3-TTS synthesis}} \\[1pt]
\multicolumn{2}{@{}l}{Pretrained base} & 5.087 & 4.886 & -- & -- \\
\cmidrule(lr){1-6}
\multirow{2}{*}{DPO} & Direct & \llloss{5.120} & \llloss{5.425} & 50.5 & 98.8 \\
 & Bridge & \llloss{5.121} & \llloss{5.381} & 56.0 & 87.0 \\
\multirow{2}{*}{WPO} & Direct & \llloss{5.114} & \llloss{5.406} & 53.5 & 99.3 \\
 & Bridge & \llgain{5.059} & \llloss{5.131} & 71.2 & 84.0 \\
\multirow{2}{*}{APO-zero} & Direct & \llgain{4.852} & \llloss{5.173} & 99.8 & 84.1 \\
 & Bridge & \llgain{4.935} & \llloss{5.068} & 95.4 & 64.2 \\
\multirow{2}{*}{NCA} & Direct & \llgain{5.073} & \llloss{5.094} & 65.5 & 96.0 \\
 & Bridge & \llgain{5.076} & \llloss{4.976} & 64.9 & 75.4 \\
\bottomrule
\end{tabular*}
\endgroup
\end{table}

%% file: Appendix/Tables/weights.tex
\begin{table}[t]
    \centering
    \small
    \setlength{\tabcolsep}{4pt}
    \renewcommand{\arraystretch}{1.08}
    \caption{
        Ablation of policy-support weighting with fixed preference
        targets and training budgets.
        WER is macro-averaged over utterances and reported in percent.
    }
    \label{tab:support_weighting}
    \begin{tabular}{@{}clrrr@{}}
        \toprule
        Objective & Weighting
        & WER$\downarrow$ & SIM$\uparrow$ & NISQA$\uparrow$ \\
        \midrule
        \multirow{3}{*}{DPO}
        & Constant  & 4.1 & 0.712 & 4.468 \\
        & Linear    & 4.1 & 0.710 & 4.464 \\
        & Parabolic & 4.3 & 0.708 & 4.458 \\
        \midrule
        \multirow{3}{*}{APO-zero}
        & Constant  & 3.6 & 0.722 & 4.505 \\
        & Linear    & 3.7 & 0.720 & 4.498 \\
        & Parabolic & 3.6 & 0.721 & 4.496 \\
        \bottomrule
    \end{tabular}
\end{table}

%% file: Appendix/Tables/phase_analysis.tex
\begin{table*}[!t]
\centering
\caption{CosyVoice checkpoints under Qwen3-TTS supervision.
    NLL entries show before-to-after measurements on
    each stage's chosen/rejected pairs.
    Pairs and stage-start diagnostic baselines differ
    across stages; the training reference remains the
    original pretrained student.
    Gain/suppression reports the percentages of
    sequences whose likelihood increases/decreases
    within that stage.
    Phases~1 and~2 use Bridge and full-expert targets,
    respectively.}
\label{tab:cosyvoice_qwen_checkpoints}
\begingroup
\footnotesize
\setlength{\tabcolsep}{3.3pt}
\renewcommand{\arraystretch}{0.98}
\begin{tabular*}{\textwidth}{@{\extracolsep{\fill}}ccccccccc@{}}
\toprule
\makecell[c]{Objective}
& \makecell[c]{Stage}
& \makecell[c]{WER(\%)$\downarrow$}
& \makecell[c]{SIM$\uparrow$}
& \makecell[c]{NISQA$\uparrow$}
& \makecell[c]{Chosen NLL}
& \makecell[c]{Rejected NLL}
& \makecell[c]{Chosen\\gain (\%)}
& \makecell[c]{Rejected\\suppression (\%)} \\
\midrule

Pretrained base & Reference
& 3.91 & 0.730 & 4.482 & -- & 2.527 & -- & -- \\
Qwen3-TTS & Expert
& 3.08 & 0.826 & 4.500 & 2.263 & -- & -- & -- \\
\midrule

\multirow{3}{*}{DPO}
& Phase 1 & 3.19 & 0.721 & 4.522
& \nllsuppress{2.227}{2.391}
& \nllsuppress{2.517}{3.292} & 2.7 & 100.0 \\
& Phase 2 & 4.23 & 0.723 & 4.498
& \nllsuppress{2.423}{2.427}
& \nllsuppress{2.759}{3.301} & 47.4 & 100.0 \\
& Direct & 3.30 & 0.713 & 4.542
& \nllsuppress{2.263}{2.581}
& \nllsuppress{2.527}{3.927} & 0.8 & 100.0 \\
\addlinespace[1pt]

\multirow{3}{*}{WPO}
& Phase 1 & 3.31 & 0.721 & 4.517
& \nllsuppress{2.227}{2.348}
& \nllsuppress{2.517}{3.113} & 3.3 & 100.0 \\
& Phase 2 & 4.33 & 0.724 & 4.504
& \nllsuppress{2.378}{2.390}
& \nllsuppress{2.682}{3.127} & 43.1 & 100.0 \\
& Direct & 3.25 & 0.714 & 4.542
& \nllsuppress{2.263}{2.533}
& \nllsuppress{2.527}{3.644} & 1.2 & 100.0 \\
\addlinespace[1pt]

\multirow{3}{*}{APO-zero}
& Phase 1 & 3.39 & 0.727 & 4.539
& \nllsupport{2.227}{2.122}
& \nllsuppress{2.517}{2.878} & 99.0 & 99.9 \\
& Phase 2 & 3.38 & 0.725 & 4.544
& \nllsupport{2.158}{2.113}
& \nllsuppress{2.627}{2.987} & 98.8 & 100.0 \\
& Direct & 3.31 & 0.722 & 4.548
& \nllsupport{2.263}{2.111}
& \nllsuppress{2.527}{3.089} & 99.9 & 100.0 \\
\addlinespace[1pt]

\multirow{3}{*}{NCA}
& Phase 1 & 3.48 & 0.730 & 4.508
& \nllsupport{2.227}{2.193}
& \nllsuppress{2.517}{2.760} & 98.5 & 100.0 \\
& Phase 2 & 3.49 & 0.728 & 4.509
& \nllsupport{2.226}{2.224}
& \nllsuppress{2.665}{2.764} & 50.7 & 99.9 \\
& Direct & 3.65 & 0.729 & 4.518
& \nllsupport{2.263}{2.225}
& \nllsuppress{2.527}{2.868} & 95.5 & 100.0 \\
\bottomrule
\end{tabular*}
\endgroup
\end{table*}